\documentclass[preprint,pre,preprintnumbers,amsmath,amssymb,showpacs,floatfix]{revtex4}
\usepackage{graphicx,bm,amsmath}

\def\<{\langle}
\def\>{\rangle}

\begin{document}
\title{Finite-size scaling in 
two-dimensional Ising spin glass models}
\author{Francesco Parisen Toldin}
\email{parisen@pks.mpg.de}
\affiliation{Max-Planck-Institut f\"ur Physik komplexer Systeme, 
N\"othnitzer Str.~38, D-01187 Dresden, Germany}
\author{Andrea Pelissetto}
\email{Andrea.Pelissetto@roma1.infn.it}
\affiliation{
Dipartimento di Fisica dell'Universit\`a di Roma ``La Sapienza'' and INFN,
Piazzale Aldo Moro 2,
I-00185 Roma,
Italy
}
\author{Ettore Vicari}
\email{vicari@df.unipi.it}
\affiliation{
Dipartimento di Fisica dell'Universit\`a di Pisa and INFN,
Largo Pontecorvo 3,
I-56127 Pisa,
Italy
}

\begin{abstract}
  We study the finite-size behavior of two-dimensional spin-glass
  models.  We consider the $\pm J$ model for two different values of
  the probability of the antiferromagnetic bonds and the model with
  Gaussian distributed couplings. The analysis of
  renormalization-group invariant quantities, the overlap
  susceptibility, and the two-point correlation function confirms that they
  belong to the same universality class. We analyze in detail the
  standard finite-size scaling limit in terms of $TL^{1/\nu}$ in the
  $\pm J$ model.  We find that it holds asymptotically. This result is
  consistent with the low-temperature crossover scenario in which
  the crossover temperature, which separates the universal high-temperature 
  region from the discrete low-temperature regime,
  scales as $T_c(L)\sim L^{-{\theta_S}}$ with $\theta_S\approx 0.5$.
\end{abstract}


\maketitle

\section{Introduction}
\label{sec:intro}

The two-dimensional (2D) Edwards-Anderson spin-glass model
\cite{EA-75} has been extensively studied in recent years in order to
investigate the interplay of disorder and frustration in 2D systems.
If frustration is sufficiently large, these systems are paramagnetic
at any finite temperature $T$.  A critical glassy behavior is only
observed for $T\to 0$. The zero-temperature behavior has been
extensively studied. It is now well understood that it depends on the
behavior of the low-energy spectrum. One should indeed distinguish
systems with a discrete energy spectrum (DES), such as the $\pm J$ Ising
model with a bimodal coupling distribution, from systems with a
continuous energy spectrum (CES), such as the Ising model with Gaussian
distributed couplings \cite{AMMP-03,JLMM-06,LMMS-06}. At $T=0$, these
two classes of systems behave quite differently. For instance, in DES
systems, the stiffness exponent vanishes, while in CES systems, we have
$\theta < 0$; recent numerical studies \cite{footnote-theta} give 
$\theta \approx -0.28$.

For finite values of $T$, CES systems show a standard critical
behavior, consistent with what is observed at $T = 0$. In particular,
one expects $\nu = -1/\theta$ and $\eta = 0$, two predictions which
are consistent with numerical data \cite{JLMM-06,KLC-07}. The behavior
of DES systems is instead more complex. In a finite box of linear size
$L$, one observes two different behaviors, which depend on how large
$L$ is compared to a temperature-dependent crossover length $L_c(T)$;
see Refs.~\cite{JLMM-06,THM-11a,THM-11b,JK-11,PPV-10} and references
therein.  For $L< L_c(T)$, the critical behavior is analogous to that
observed at $T=0$. The system shows an effective long-range spin-glass
order and its critical behavior can be predicted using droplet
theory \cite{THM-11a,JK-11}.  On the other hand, for $L>L_c(T)$, the
system is effectively paramagnetic.  Equivalently, at fixed $L$, one
observes the two regimes for $T<T_c(L)$ and $T>T_c(L)$, respectively,
where $T_c(L)$ is the corresponding crossover temperature. Note that
this discrete behavior can only be observed for finite values of $L$,
since $T_c(L)\to 0$ for $L\to\infty$. 
Of course, since $T_c(L)$ is an effective finite-size temperature, 
the crossover temperature is not uniquely defined, 
and many different definitions can be used.
One of the basic questions is whether the critical behavior
of DES systems for $T>T_c(L)$ is the same as that observed in CES
systems. The numerical results of Ref.~\cite{JLMM-06} strongly
suggested that this is the case.  However, those conclusions were
later questioned in Ref.~\cite{KLC-07}, on the basis that much larger
lattices were needed to show it conclusively.

In a renormalization-group (RG) picture, the two regimes can be
interpreted as due to two different fixed points (FPs)
\cite{JLMM-06,JK-11}: a stable FP---the same that controls the
critical behavior of CES systems---which describes the infinite-volume
behavior up to $T=0$, and an unstable FP, present only in DES systems,
which controls the low-temperature behavior for $T < T_c(L)$.

In order to fully specify the two regimes, one should predict how
$T_c(L)$ scales with the size $L$.  A free-energy argument, based on the
energy difference and degeneracies of the two lowest-energy states~\cite{KLC-07}, suggests $T_c(L)\sim 1/\ln L$ as $L\to
\infty$. However, recently, using droplet theory,
Refs.~\cite{THM-11a,THM-11b,JK-11} suggested a power-law behavior
\begin{equation}
L_c(T)\sim T^{-1/\theta_S}, \qquad\qquad
T_c(L) \sim L^{-\theta_S}.
\label{TcL}
\end{equation}
Reference \cite{THM-11a} predicted $\theta_S \approx 0.50(1)$,
which appears to be consistent with their numerical data for the 
twist free energy \cite{THM-11a} and the two-point correlation function
\cite{THM-11b}, as well as with previous results \cite{SK-94,LMMS-06}. 
A calculation in a hierarchical model \cite{JK-11} gives 
a similar result of $\theta_S \approx  0.37$. 
These calculations indicate that although $\theta_S$ is quite small,
it is nonetheless larger than the exponent $1/\nu = -\theta \approx 0.28 $ 
($\theta$ is the stiffness exponent in CES systems).

In this paper, we investigate again the question of universality, by
comparing the finite-size scaling (FSS) of the $\pm J$ model for two
values of the disorder parameter $p=0.5,\,0.8$ and the model with
Gaussian distributed couplings (henceforth we call it the Gaussian model).
The FSS analysis in terms of RG invariant quantities (for example, the
plots of the Binder cumulants versus the ratio $\xi/L$, with all quantities
being defined in terms of the overlap variables) shows that the two
models belong to the same universality class, confirming the
conclusions of Refs.~\cite{JLMM-06,LMMS-06}. Indeed, the $\pm J$ data
have the same FSS behavior as the Gaussian data, if we only consider
the $\pm J$ model results corresponding to temperatures larger than
the crossover temperature.  Then, we focus on the validity of the
standard FSS in terms of the variable $TL^{1/\nu}$, which is a rather
subtle point in DES systems.  Standard FSS exists only if
$T_c(L)L^{1/\nu}\to 0$ for $L\to \infty$.  If we assume $T_c(L)\sim
L^{-\theta_S}$, then since $T_c(L)L^{1/\nu} \sim L^{1/\nu - \theta_S}$, FSS
can be observed only if $\theta_S > 1/\nu\approx 0.28$.  This implies
that if $T_c(L) \sim 1/\ln L$
\cite{KLC-07,PPV-10}, the FSS limit $T\to 0$, $L\to\infty$ at
fixed $TL^{1/\nu}$ does not exist in DES systems.  On the other hand,
if Eq.~(\ref{TcL}) holds with $\theta_S \approx 0.50$, FSS holds also
in DES models.  However, the approach to the asymptotic limit is quite
slow. The region in which no FSS is observed, which corresponds to
$TL^{1/\nu} \lesssim T_c(L)L^{1/\nu}$, shrinks slowly, as $L^{1/\nu -
\theta_S}\sim L^{-0.2}$.  The comparison of the Monte Carlo (MC)
simulations of the $\pm J$ and Gaussian models shows quite
convincingly that for fixed $TL^{1/\nu}$ close to $T_c(L) L^{1/\nu}$,
the $\pm J$ model data converge toward the data of the Gaussian
model, confirming the existence of the standard FSS, hence the
power-law behavior (\ref{TcL}) with $\theta_S > 1/\nu$. A reanalysis
of the freezing temperature $T_f(L)$, defined in Ref.~\cite{PPV-10}
from the freezing of $\xi/L$ and of the Binder cumulants [at fixed $L$, 
they are approximately constant for $T<T_f(L)$], is consistent with
Eq.~(\ref{TcL}) with $\theta_S \approx 0.4$, which is close to the
estimate of Ref.~\cite{THM-11a}. The freezing temperature $T_f(L)$
should represent a correct effective definition for $T_c(L)$, although
deviations from the universal FSS behavior are expected for somewhat
larger values of $T$.

We also investigate the FSS behavior of the magnetization and the
two-point correlation function of the overlap variables. We find that
the data are consistent with the hyperscaling relation $2\beta=\eta\nu$. 
However, the data are not sufficiently precise to provide a precise
determination of $\eta$, being consistent with a small value of
$\eta\lesssim 0.2$, including $\eta=0$.  In Ref.~\cite{THM-11b}, the
authors showed that a properly subtracted overlap correlation function
scales in the temperature region they consider, which essentially
corresponds to $T\lesssim T_c(L)$. Here we consider the opposite regime, $T >
T_c(L)$.  We find that standard FSS as well as universality hold for
the overlap correlation function.

The paper is organized as follows. In Sec.~\ref{sec:mc}, we define 
the models and the quantities we investigate. Section \ref{sec:results}
reports the numerical results of our FSS analysis: in
Sec.~\ref{sec3.1}, we discuss the RG invariant couplings, such as the
ratio $\xi/L$ and the cumulants of the overlap variable, focusing
mainly on the question of the validity of FSS in terms of
$TL^{1/\nu}$; in Sec.~\ref{sec3.2}, we discuss the overlap
magnetization and susceptibility; and finally, in Sec.~\ref{sec3.3},
we discuss the two-point correlation function. In Sec.~\ref{sec4}, we
present our conclusions.

\section{Models and definitions}
\label{sec:mc}

We consider the 2D Ising model on a square lattice with Hamiltonian
\begin{equation}
{\cal H} = - \sum_{\langle xy \rangle} J_{xy} \sigma_x \sigma_y,
\label{lH}
\end{equation}
where $\sigma_x=\pm 1$, the sum is over all pairs of lattice
nearest-neighbor sites, and the exchange interactions $J_{xy}$ are
uncorrelated quenched random variables.  We consider a model
with Gaussian bond distribution,
\begin{equation}
P(J_{xy}) \sim \exp (-J_{xy}^2/2)
\label{gaudistr}
\end{equation}
(in the following, we call it the Gaussian model).
We also consider the $\pm J$ model where the
couplings $J_{xy}$ take values $\pm J$ with probability distribution
\begin{equation}
P(J_{xy}) = p \delta(J_{xy} - J) + (1-p) \delta(J_{xy} + J).
\label{probdis}
\end{equation}
As in Ref.~\cite{PPV-10}, we consider $p=0.5$ and $0.8$.  We recall
that for sufficiently large frustration, i.e., 
for $0.11 \lesssim p \lesssim 0.89$,
the model shows a zero-temperature glassy critical behavior, with 
a paramagnetic low-temperature phase.
Ferromagnetism can only be observed 
for $p>p^*=0.89093(3)$ \cite{PPV-09}.

The critical modes at the glassy transition are those related to the overlap
variable $q_x \equiv \sigma_x^{(1)} \sigma_x^{(2)}$, where the spins
$\sigma_x^{(i)}$ belong to two independent replicas with the same disorder
realization $\{J_{xy}\}$.  In our Monte Carlo (MC) simulations, we compute the 
overlap magnetization 
\begin{equation}
  m = \frac{1}{L^2} [\langle|\sum_x q_x|\rangle],
\end{equation}
the overlap susceptibility $\chi$, and the
second-moment correlation length $\xi$ defined from the correlation
function
\begin{equation}
G_o(x) \equiv [\langle q_0 \,q_x \rangle] =
[\langle \sigma_0 \,\sigma_x \rangle^2], 
\label{go}
\end{equation}
where the angular and the square brackets indicate the thermal average
and the quenched average over disorder, respectively. We define $\chi
\equiv \sum_{x} G_o(x)$ and
\begin{eqnarray}
\xi^2 \equiv  \frac{1}{4 \sin^2 (p_{\rm min}/2)} 
\frac{\widetilde{G}_o(0) - \widetilde{G}_o(p)}{\widetilde{G}_o(p)},
\label{xidefffxy}
\end{eqnarray}
where $p = (p_{\rm min},0)$, $p_{\rm min} \equiv 2 \pi/L$, and
$\widetilde{G}_o(q)$ is the Fourier transform of $G_o(x)$.  We also
consider some quantities that are invariant under RG transformations
in the critical limit, which we call phenomenological couplings.  We
consider the ratio $\xi/L$ and the quartic cumulants
\begin{equation}
U_{4}  \equiv \frac{[ \rho_4 ]}{[\rho_2]^{2}}, \quad
U_{22} \equiv  \frac{[ \rho_2^2 ]-[\rho_2]^2}{[\rho_2]^2},
\label{uc}
\end{equation}
where $\rho_{k} \equiv \langle \; ( \sum_x q_x\; )^k \rangle$.  

In the case of a $T=0$ transition with a nondegenerate ground state,
as expected in CES systems, the overlap magnetization
exponent $\beta$  vanishes, and 
$U_{4}\to 1$ and $U_{22}\to 0$ for $T\to 0$. Moreover,
assuming the hyperscaling relation $2\beta = \eta\nu$, we obtain $\eta=0$, thus
$\chi\sim \xi^2$ for $T\to 0$.

\section{Finite-size scaling behavior}
\label{sec:results}

In order to study the FSS behavior, we extend the MC simulations of
the $\pm J$ model at $p=0.5$ and $0.8$ presented in
Ref.~\cite{PPV-10}; we perform further simulations of the $\pm J$
Ising model at $p=0.8$ on finite square lattices of sizes $L=16,32$,
and of the Gaussian model for $L = 8,12, 16$.  We use the Metropolis
algorithm and the random-exchange method~\cite{raex}.  For the
Gaussian model, we study the temperature interval $T_{\rm min}\le T
\lesssim 1.6$, with $T_{\rm min}=0.2$ ($L=8$) and $T_{\rm min}=0.167$
($L=12,16$). We average over a large number of disorder samples, i.e., 
$10^4$ for each $T$ and $p$.

\subsection{RG invariant couplings} \label{sec3.1}

\begin{figure}
\includegraphics[width=17em,keepaspectratio]{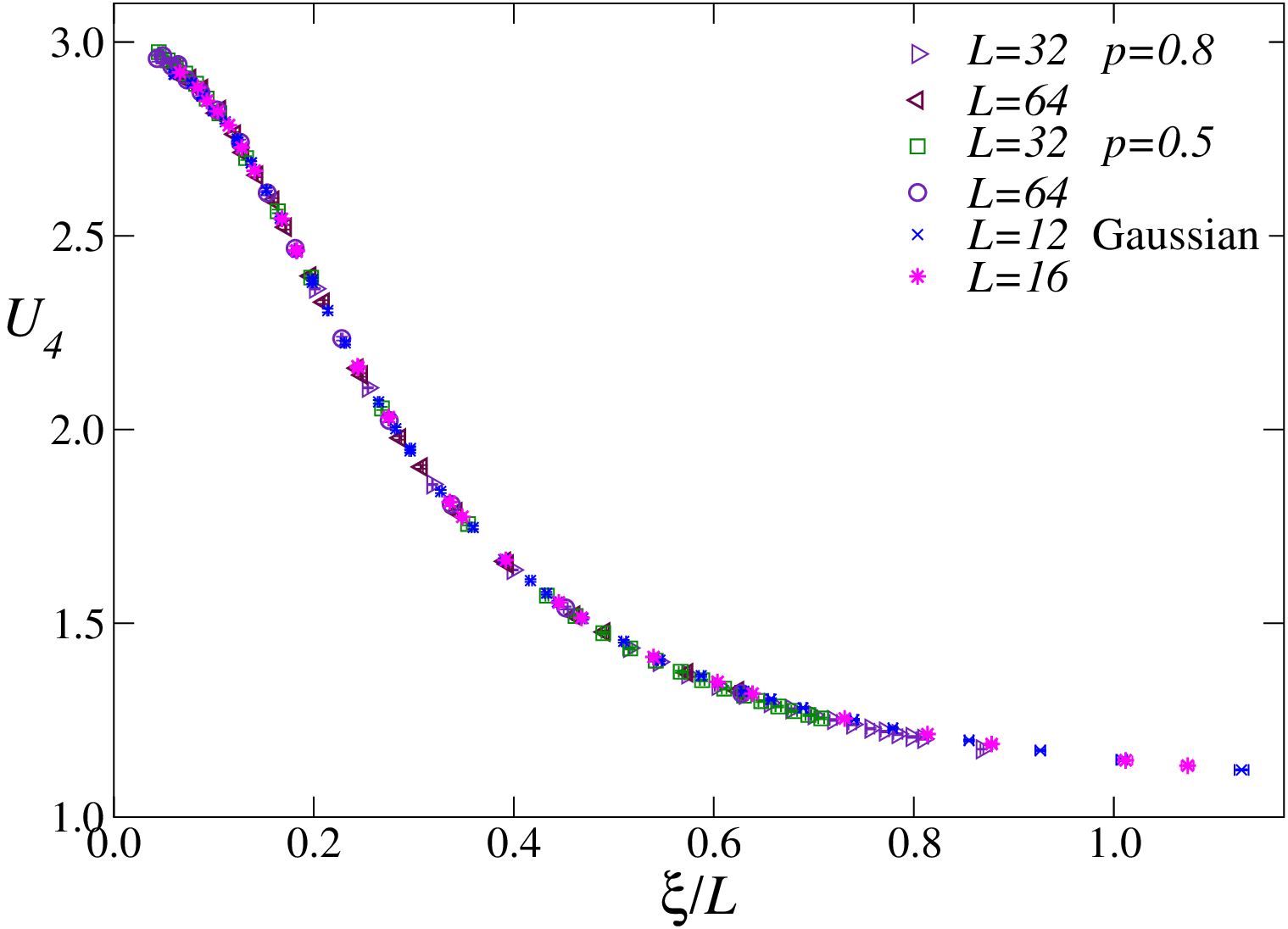} 
\hspace{1truecm}
\includegraphics[width=17em,keepaspectratio]{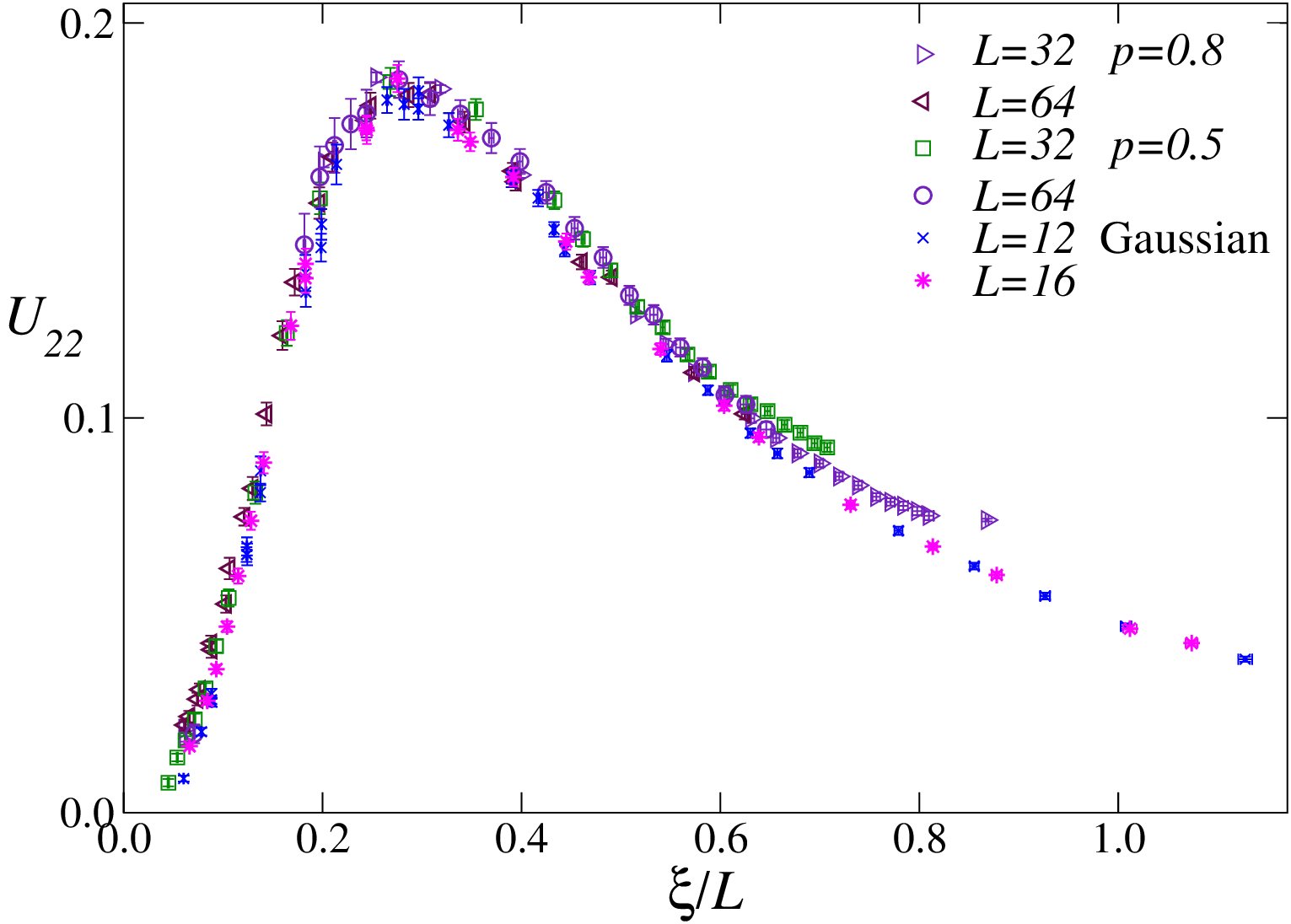}
\caption{(Color online) 
The cumulants $U_4$ and $U_{22}$ vs
$\xi/L$.  We present data for the $\pm J$ model ($p= 0.5$ and $0.8$)
and for the Gaussian model. Only the data for the largest lattices are
included for clarity.  
}
\label{FSSxiL}
\end{figure}

To begin with, we wish to check that the $\pm J$ model and the
Gaussian model belong to the same glassy universality class, extending
the FSS analyses of Refs.~\cite{JLMM-06,KLC-07,PPV-10}.  For this
purpose, we consider $U_4$ and $U_{22}$ as a function of $\xi/L$. Our
numerical results for the largest values of $L$ are reported in
Fig.~\ref{FSSxiL}. No scaling corrections are visible in the plot of
$U_4$, as already observed in Refs.~\cite{KLC-07,PPV-10}, while
slightly larger corrections appear in the case of $U_{22}$ for $\xi/L
\gtrsim 0.6$. It is, however, evident that as $L$ increases, the
differences between the Gaussian model results and those for the $\pm
J$ model decrease.  Thus, these results, together with those presented
in Refs.~\cite{JLMM-06,KLC-07} (in Ref.~\cite{JLMM-06}, other CES and
DES systems were considered), confirm that all models belong to the
same universality class.

Now we investigate the question of the existence of standard FSS as
a function of $T L^{1/\nu}$. For a RG-invariant quantity $R$, we expect
\begin{equation}
   R = h_R(x), \qquad\qquad x \equiv a T L^{1/\nu},
\label{RFSS}
\end{equation}
where $a$ is a nonuniversal constant that depends on the model
but not on the quantity $R$, which can be chosen so that $h_R(x)$ is 
model independent. 

\begin{figure}
\begin{tabular}{cc}
\includegraphics[width=18em,keepaspectratio]{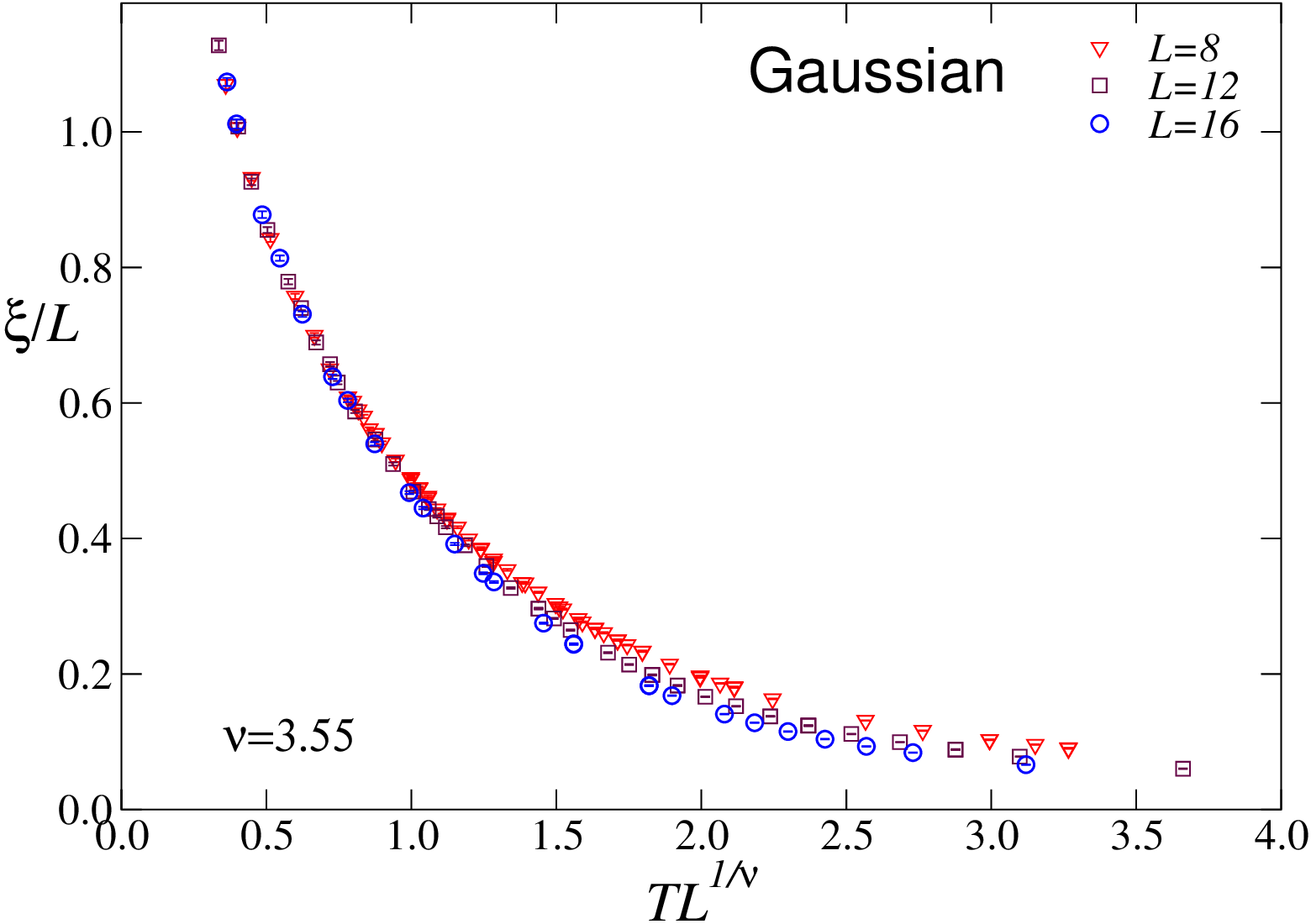} &
\includegraphics[width=18em,keepaspectratio]{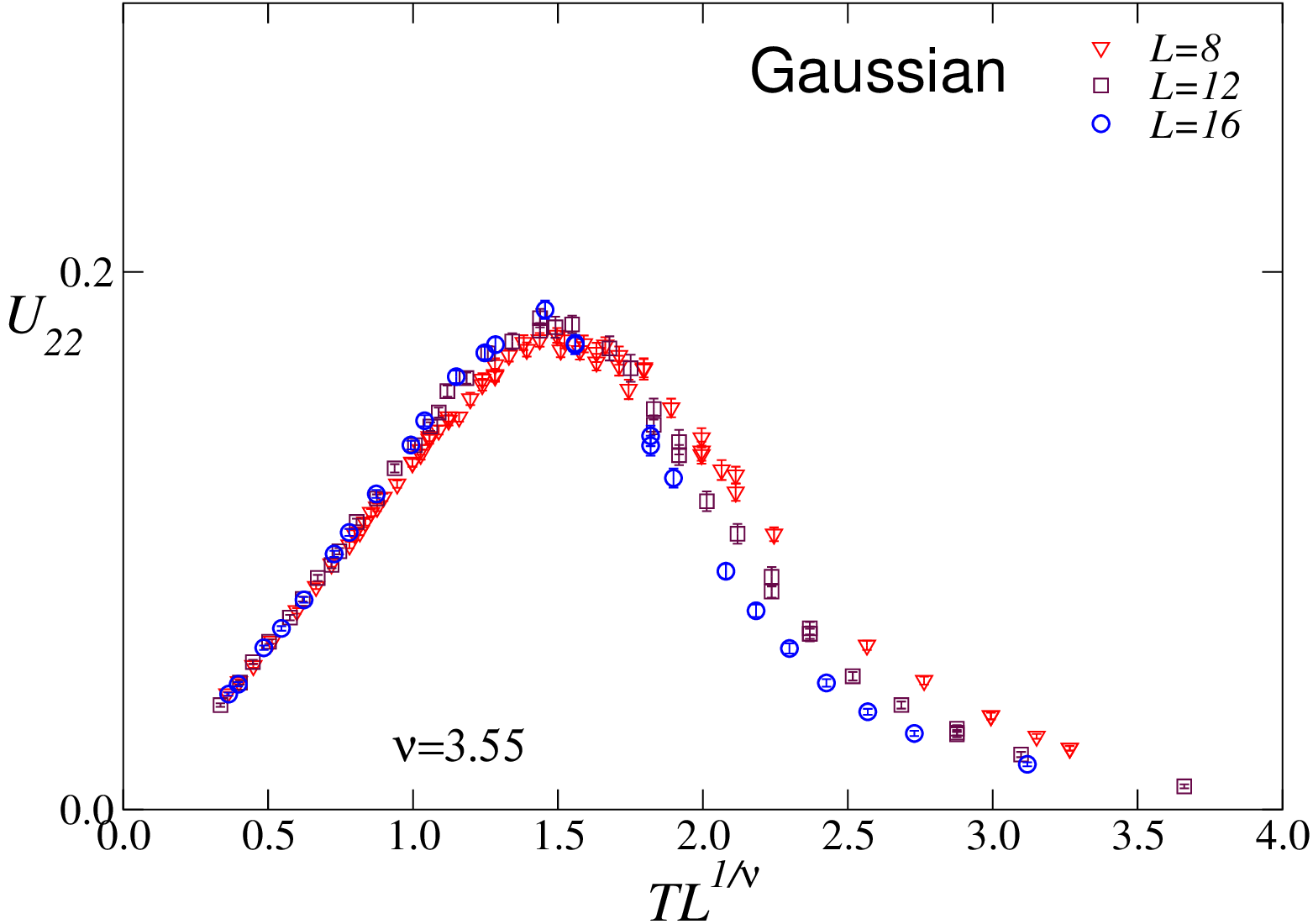} \\
\includegraphics[width=18em,keepaspectratio]{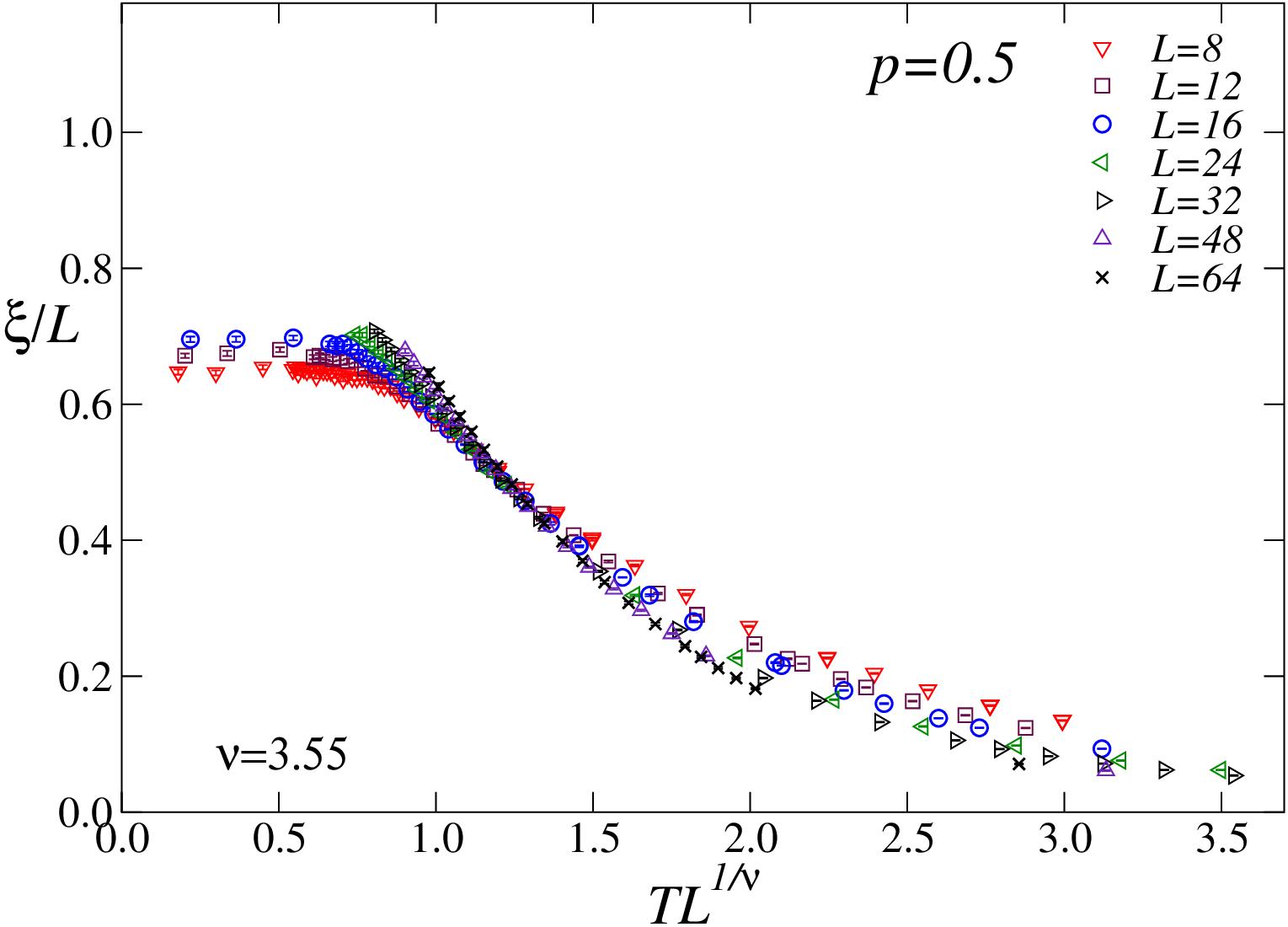} &
\includegraphics[width=18em,keepaspectratio]{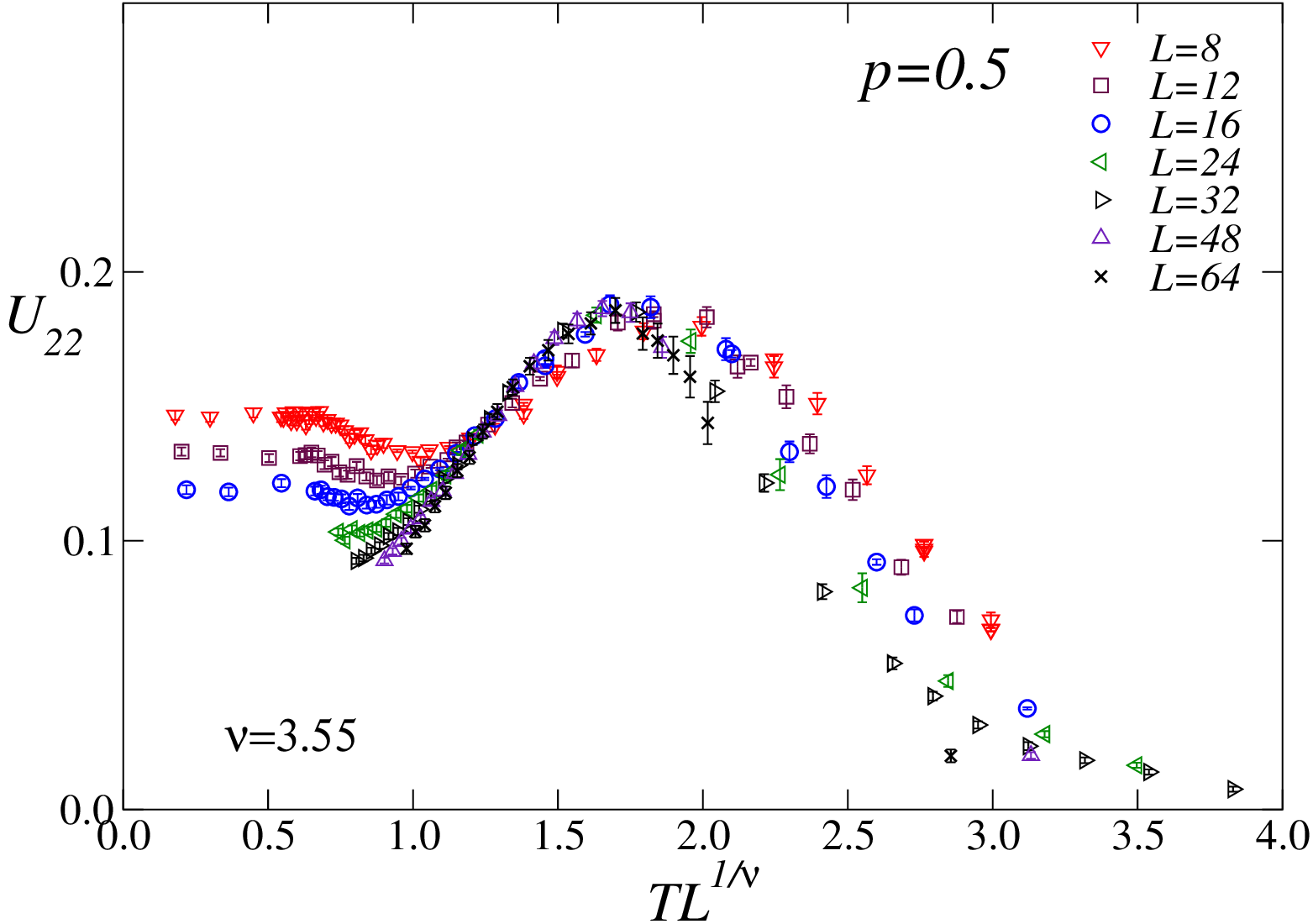} \\
\includegraphics[width=18em,keepaspectratio]{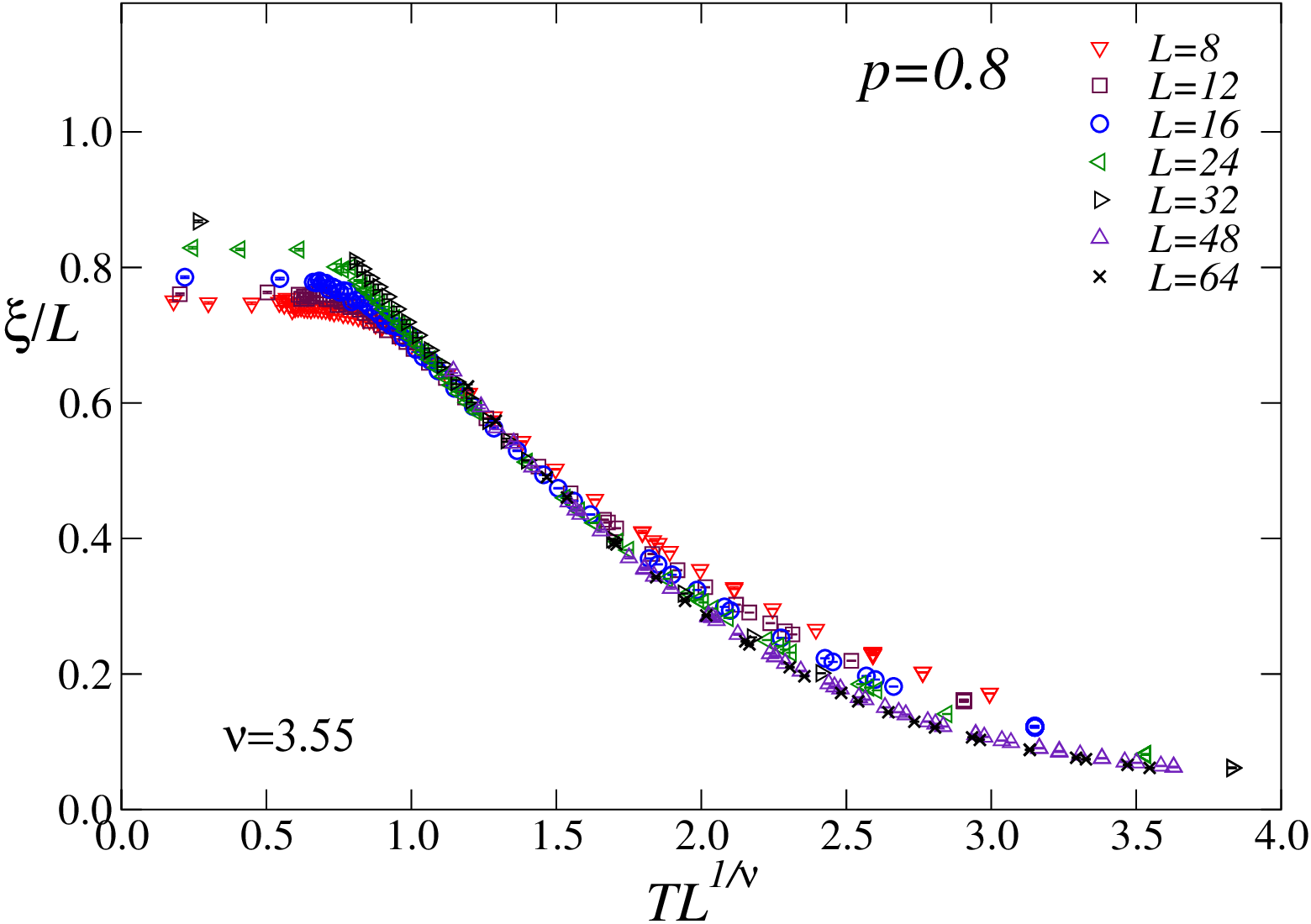} &
\includegraphics[width=18em,keepaspectratio]{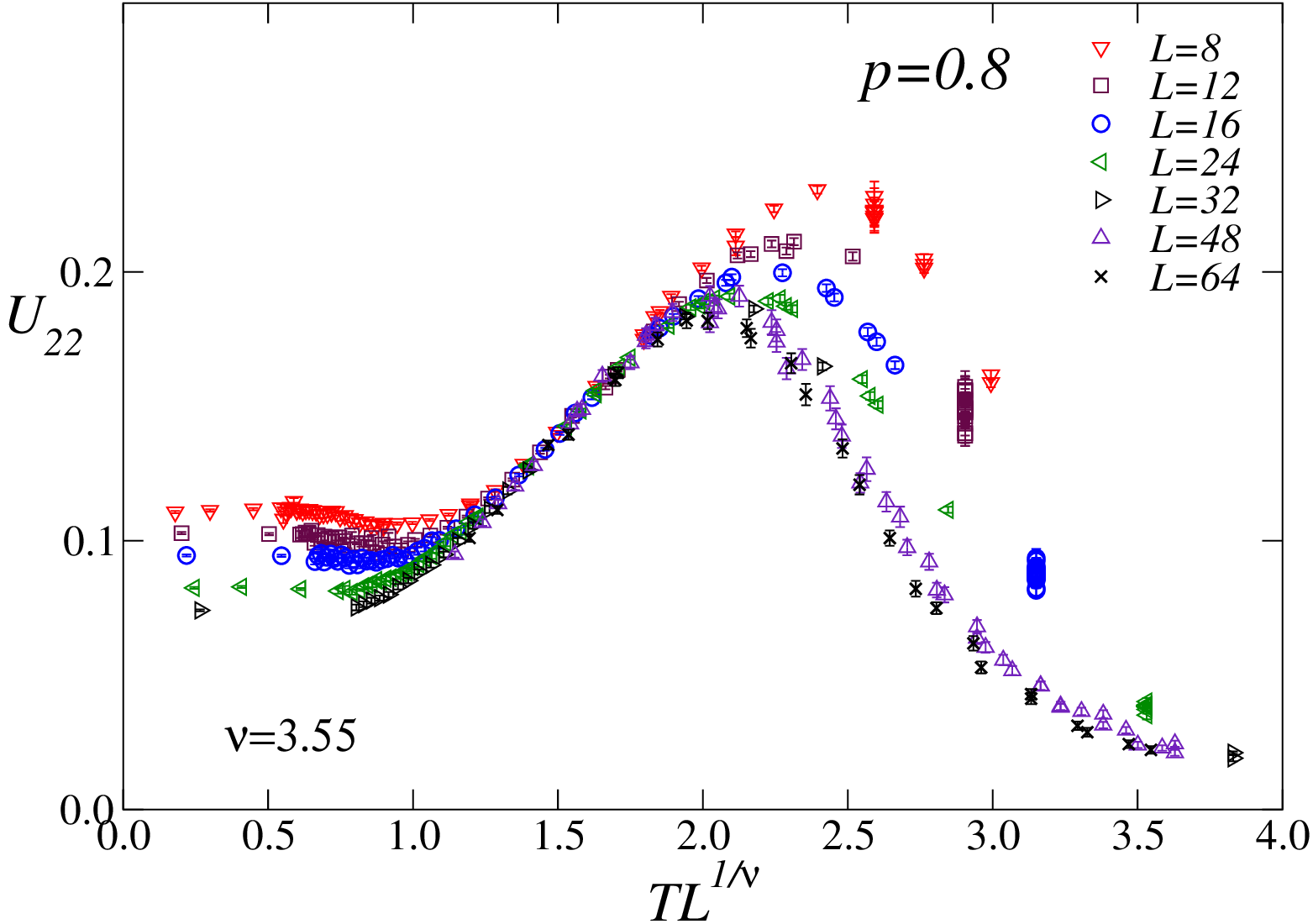} \\
\end{tabular}
\caption{(Color online) The phenomenological couplings $\xi/L$ (left)
and $U_{22}$ (right) vs $ T L^{1/\nu}$. We report the results for
the Gaussian model (top), and for the $\pm J$ model at $p= 0.5$ (middle) 
and at $0.8$ (bottom).  }
\label{FSSstd}
\end{figure}

First, we consider the data for the Gaussian model. The numerical
results for $\xi/L$ and $U_{22}$ are reported versus $T L^{1/\nu}$ in
Fig.~\ref{FSSstd} (top). We use $\nu = 3.55$, which corresponds to
$\theta = - 1/\nu = -0.282$, which in turn is the present best estimate of the
stiffness exponent of the Gaussian model \cite{footnote-theta}.  The
data show that $\xi/L$ and $U_{22}$ scale reasonably as a function of
$T L^{1/\nu}$ and clearly appear to approach a FSS limit with
increasing $L$. Scaling violations increase as $T L^{1/\nu}$
increases.  This is not unexpected since the correct scaling variable
is the combination $u_T(T) L^{1/\nu}$, where $u_T(T)$ is the nonlinear
scaling field associated with the temperature.  Considering $T
L^{1/\nu}$ as the scaling variable corresponds to expanding $u_T(T)$ to
first order in the temperature, an approximation which is expected to
work well only when $T$ is small. On the other hand, the region $TL^{1/\nu}
\gtrsim 2$ corresponds to temperatures $T\gtrsim 1$
for the lattice sizes considered here.
Of course, we cannot exclude the additional presence of
nonanalytic scaling corrections, which increase as $T L^{1/\nu}$
increases.

Then, we consider the data for the $\pm J$ model; see
Fig.~\ref{FSSstd}. The data show essentially three types of behavior,
depending on the value of $TL^{1/\nu}$.  For $TL^{1/\nu}\lesssim 1.5$, 
no scaling is observed.  This may be explained by the fact that the
data in this region are below the crossover temperature, i.e.,  they
correspond to $T<T_c(L)$, and thus are outside the regime in which
FSS is supposed to hold. Then, there is an intermediate region, $1.5
\lesssim TL^{1/\nu} \lesssim 2.0$, where data show scaling with small
corrections --- this is particularly evident for $U_{22}$. For
$TL^{1/\nu} \gtrsim 2.0$, the corrections are larger, but the results
appear to rapidly converge  to a limiting curve: for both $\xi/L$ and $U_{22}$, the
results satisfying $L\ge 32$ are very close to each other.  We
conclude that at least for $TL^{1/\nu} \gtrsim 1.5$, FSS apparently
holds.

\begin{figure}
\begin{tabular}{cc}
\includegraphics[width=18em,keepaspectratio]{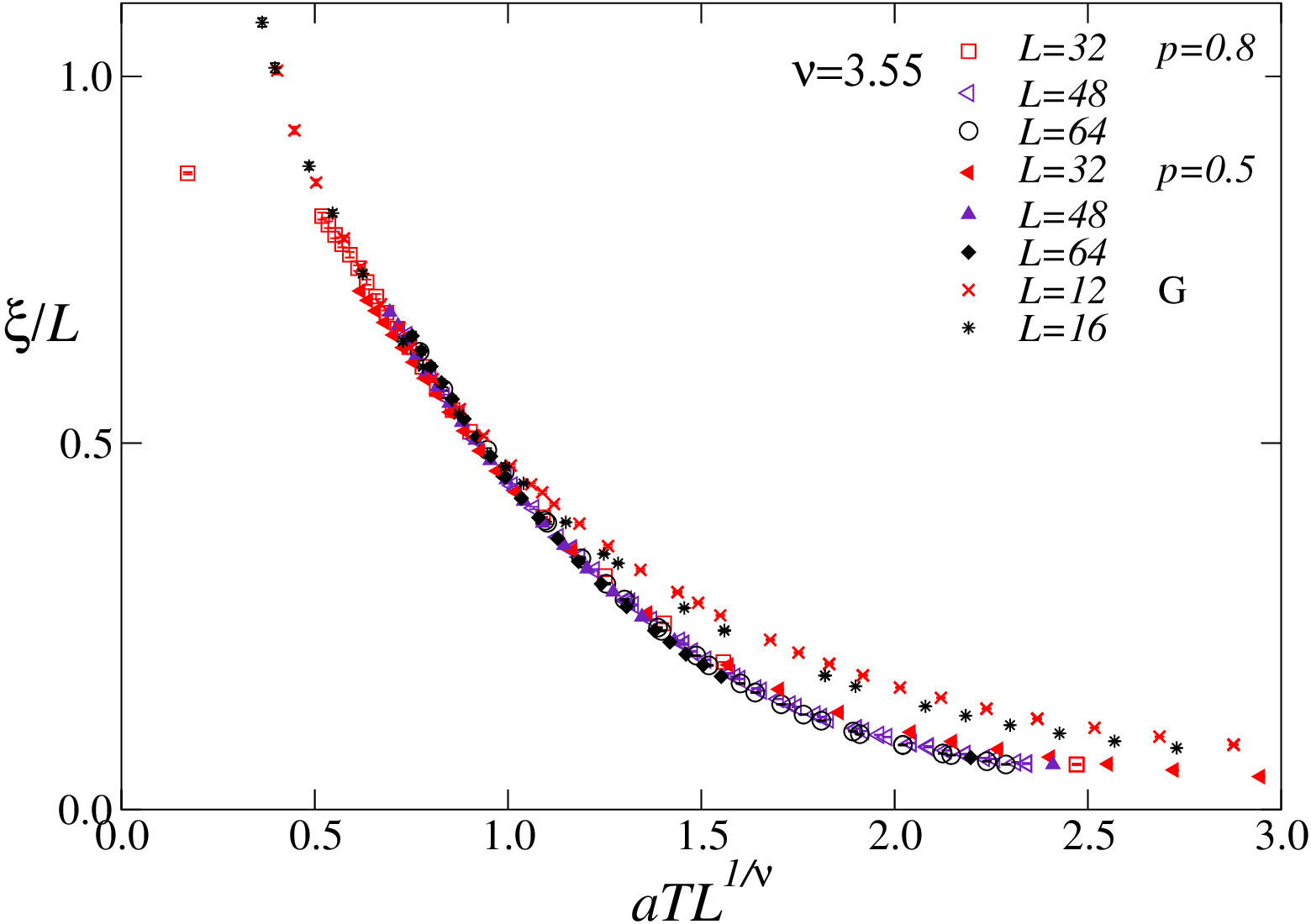} &
\includegraphics[width=18em,keepaspectratio]{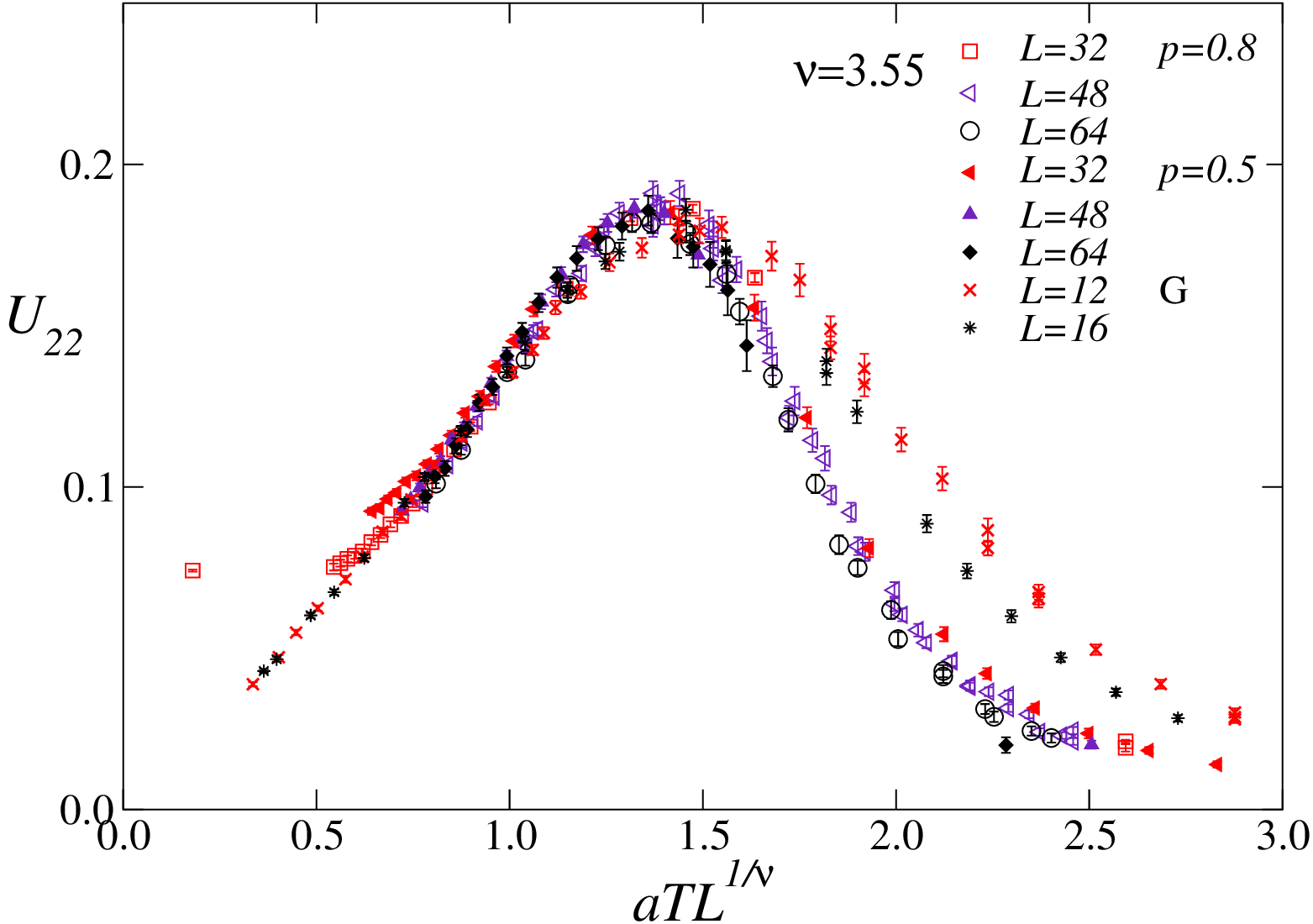} \\
\end{tabular}
\caption{(Color online) The phenomenological couplings $\xi/L$ (left)
and $U_{22}$ (right) versus $a T L^{1/\nu}$. We report the results for
the $\pm J$ model at $p=0.8$ and $0.5$ (only $L\ge 32$) and for
the Gaussian (G) model (only $L\ge 12$).  We fix $a = 1$, $1.5$, and $1.3$ for the Gaussian model, and for the $\pm J$ model at $p=0.8$ and
$0.5$, respectively.  }
\label{FSSstd2}
\end{figure}

Furthermore, we verify the universality of the FSS behavior by
comparing the results for the function $h_R(x)$ defined in
Eq.~(\ref{RFSS}). For this purpose, we should first fix the
model-dependent constant $a$ that appears in Eq.~(\ref{RFSS}).  We
determine it by requiring the FSS curves for $\xi/L$ to coincide in
the region in which $\xi/L\approx 0.5$. Indeed, in this range of
values of $\xi/L$, we observe small scaling deviations in all of the models we
consider. If we set $a = 1$ for the Gaussian model, then for the
$\pm J$ model, we obtain 
$a(p=0.5) \approx 1.3$ and $a(p = 0.8) \approx 1.5$.
In Fig.~\ref{FSSstd2}, we plot together the data for the Gaussian model and
the $\pm J$ models at $p = 0.8$ and $0.5$.  For clarity, we only
report the data with $L \ge 32 $ for the $\pm J$ model and the
results with $L \ge 12$ for the Gaussian model. With this choice, there
is only one point (it belongs to the $\pm J$ model with $p=0.8$ and
corresponds to $L=32$) which belongs to the region $T<T_c(L)$. This
point is clearly visible in the figures as an isolated point. If we
discard it, we observe good scaling up to $a T L^{1/\nu} \lesssim
1.5$: the $\pm J$ model data and the Gaussian data fall on top of each
other with good precision.  For $a T L^{1/\nu} \gtrsim 1.5$, the $\pm
J$ data scale reasonably.  The data of the Gaussian model, which correspond
to significantly smaller lattices, show significant scaling
corrections. It is, however, reassuring that the trend is correct: as
$L$ increases, they approach the $\pm J$ results.

\begin{figure}
\begin{tabular}{cc}
\includegraphics[width=18em,keepaspectratio]{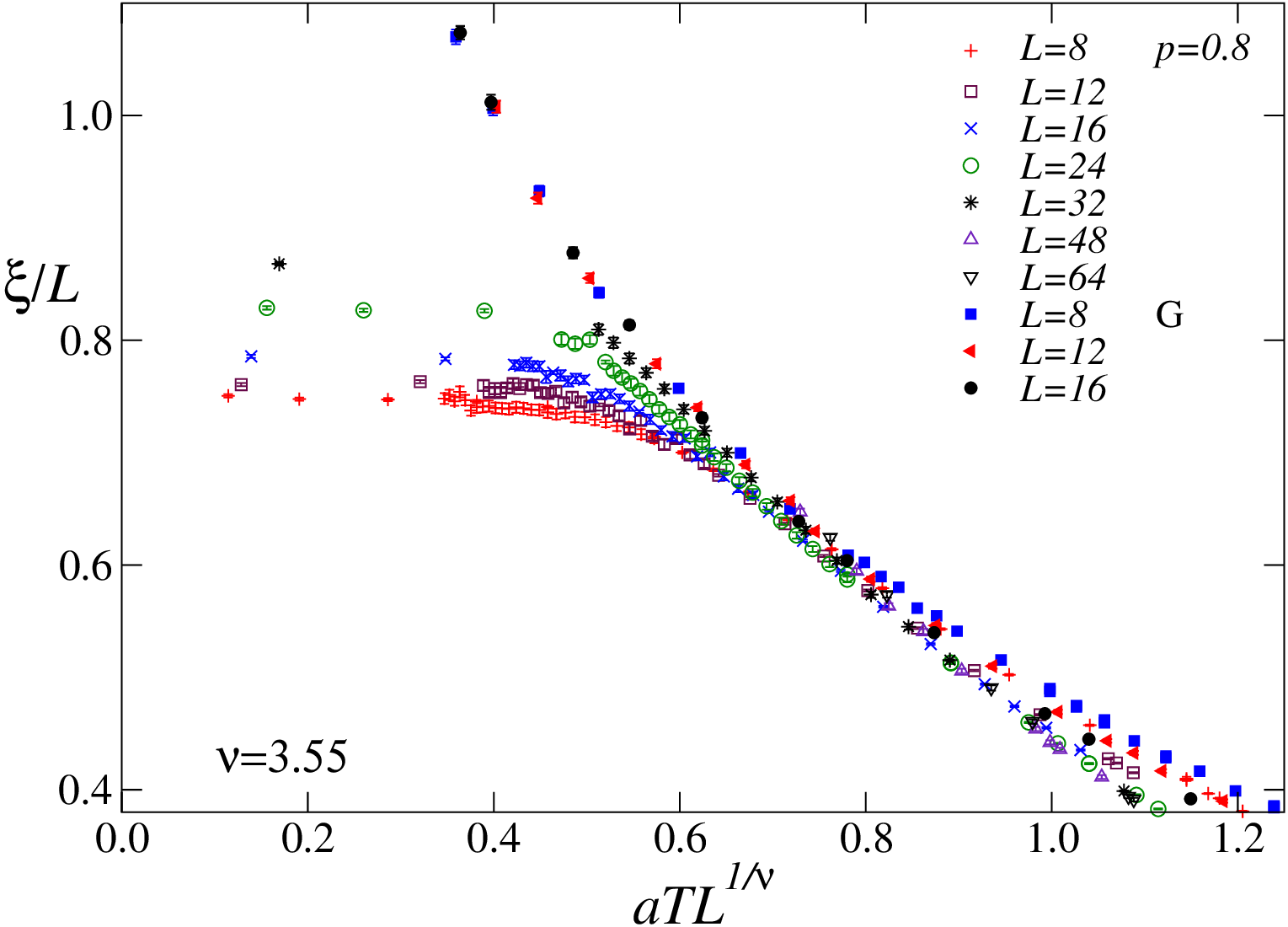} &
\includegraphics[width=18em,keepaspectratio]{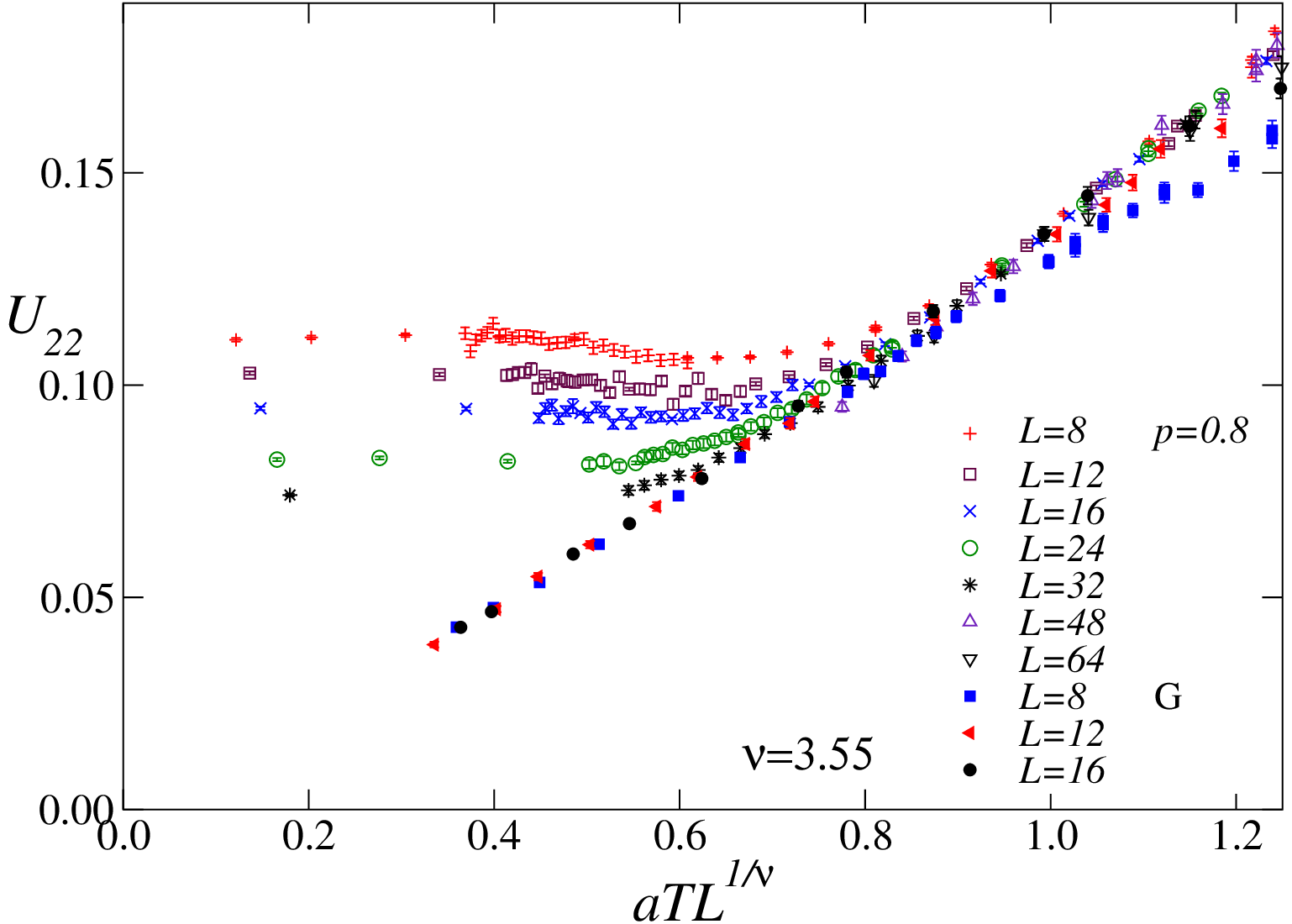} \\
\end{tabular}
\caption{(Color online) The phenomenological couplings $\xi/L$ (left)
and $U_{22}$ (right) vs $a T L^{1/\nu}$ for $a T L^{1/\nu} <
1.25$. We report the results for the $\pm J$ model at $p=0.8$ and for
the Gaussian model.  We fix $a = 1$ for the Gaussian (G) model and $a =
1.5$ for the $\pm J$ model at $p=0.8$. }
\label{FSSstd3}
\end{figure}

In order to understand the behavior close to the crossover temperature
$T_c(L)$, in Fig.~\ref{FSSstd3} we report results for all values of
$L$, but only for $a T L^{1/\nu} < 1.25$. It is clear that the
deviations between $\pm J$ and Gaussian data slowly decrease as $L$
increases (the same occurs for $p=0.5$, not shown). This is consistent with the idea that FSS in terms of $TL^{1/\nu}$ holds asymptotically, and, 
hence, with the prediction $T_c(L) \sim L^{-\theta_S}$ with $\theta_S >
1/\nu$.  The approach is, however, very slow. Indeed, the region in
which FSS does not hold is predicted to shrink as
$L^{-\theta_S+1/\nu}\sim L^{-0.2}$.

\begin{figure}
\begin{tabular}{cc}
\includegraphics[width=18em,keepaspectratio]{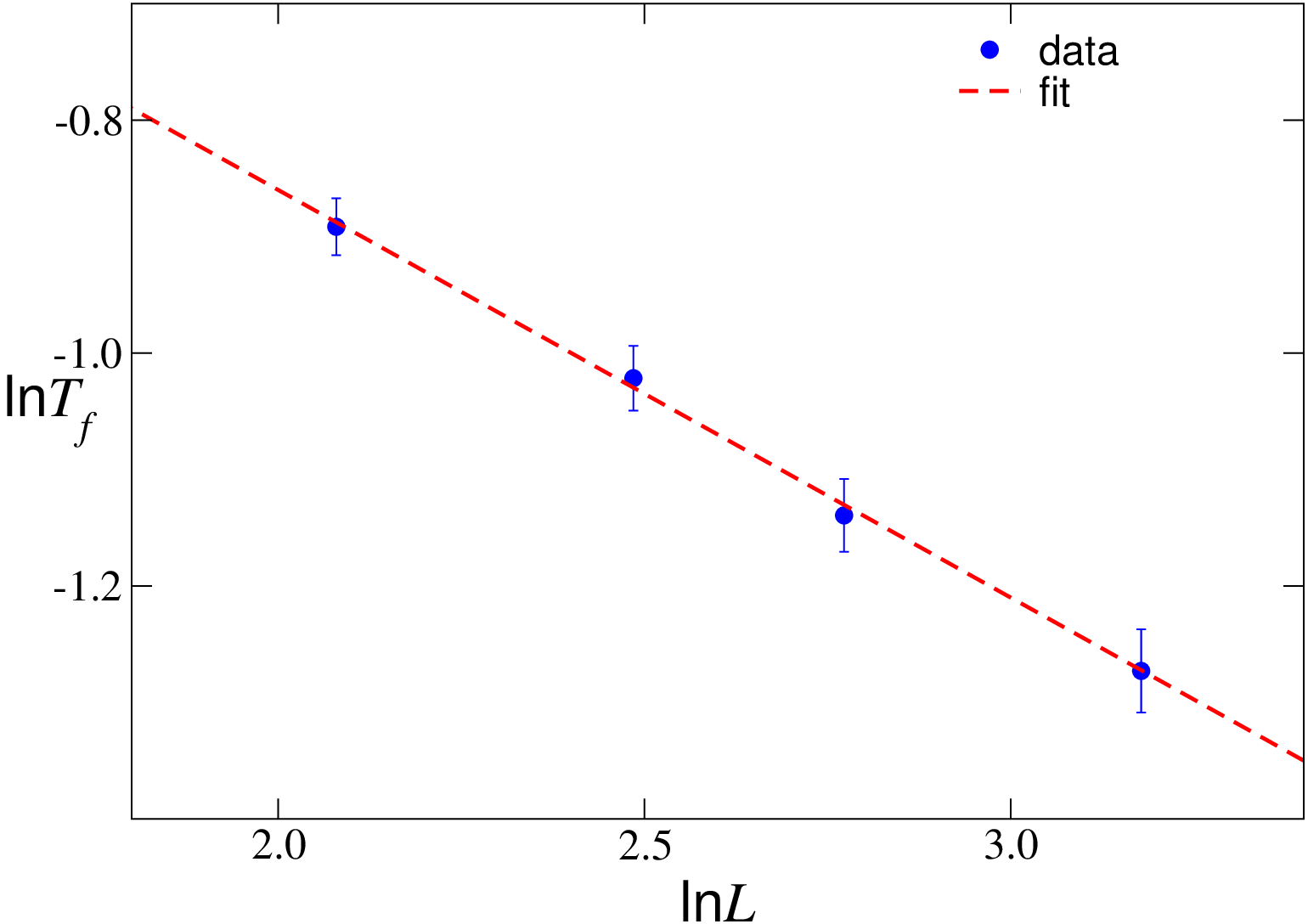} \\
\end{tabular}
\caption{(Color online) Log-log plot of the freezing temperature
$T_f(L)$ vs $L$ for the $\pm J$ model at $p=0.8$.  The dashed line
shows a fit to the data, corresponding to $T_f(L)\sim L^{-0.35}$.  }
\label{plotTc}
\end{figure}

This power-law behavior is also supported by the scaling of the
freezing temperature $T_f(L)$ defined in Ref.~\cite{PPV-10}. 
For each value of $L$, $\xi/L$ and $U_4$ become constant 
for small $T$, assuming values $(\xi/L)_f$ and $U_{4,f}$. 
Then, one defines $T_f(L)$ as the largest temperature of the
region in which $\xi/L \approx (\xi/L)_f$ and $U_4 \approx U_{4,f}$.
A fit of the data to a power-law behavior \cite{logw} gives
$T_f(L) \sim L^{-0.35}$; see Fig.~\ref{plotTc}.
Given the {\em ad hoc} procedure \cite{footnote-etf} used to determine
$T_f(L)$, it is difficult to give a reliable error for the result. It is,
however, reassuring that the estimate satisfies the bound $\theta_S >
1/\nu$ and is close to the estimates of Refs.~\cite{LMMS-06,THM-11a}.

\subsection{Overlap magnetization and susceptibility} \label{sec3.2}

\begin{figure}
\begin{tabular}{c}
\includegraphics[width=22em,keepaspectratio]{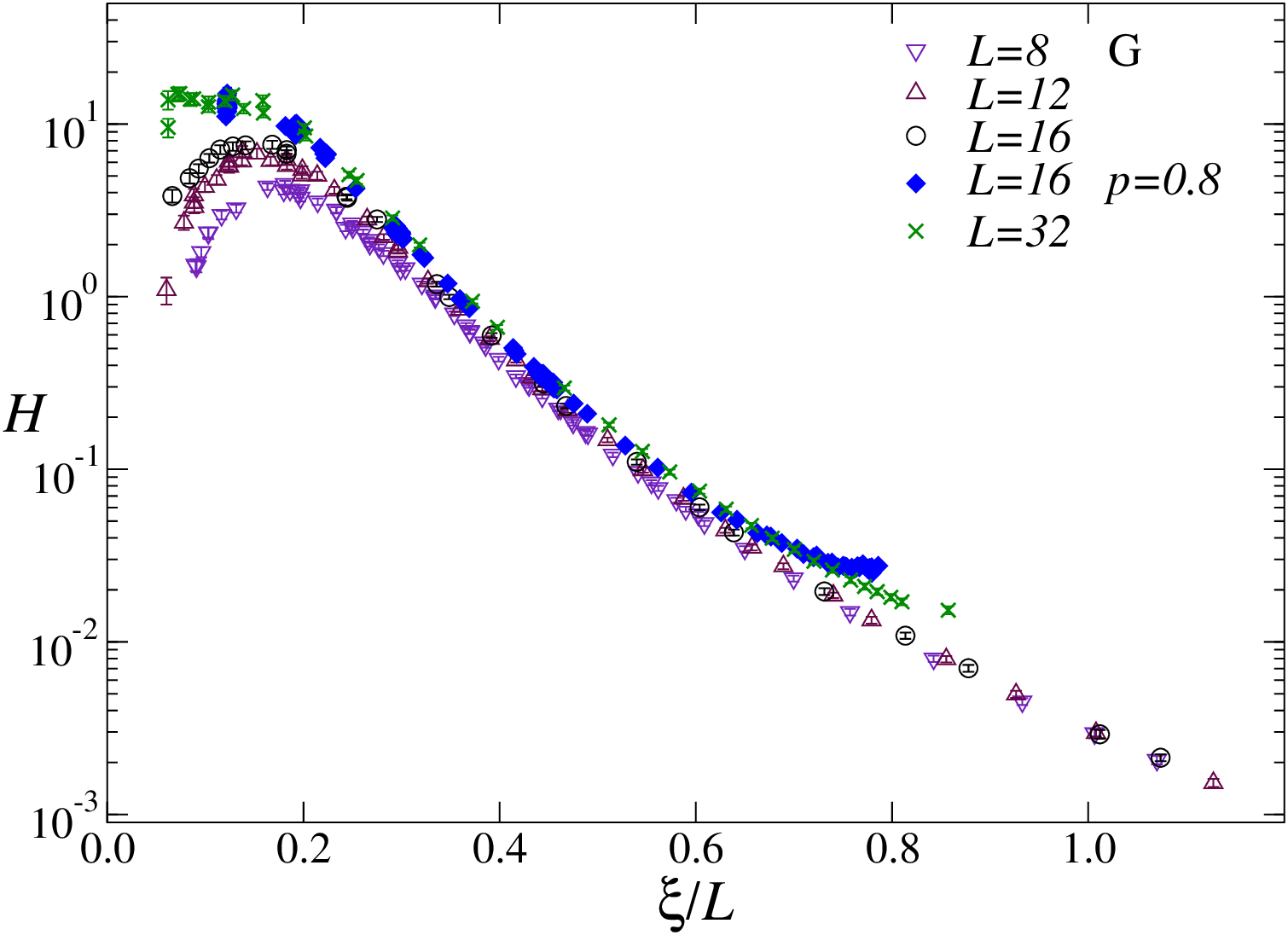} \\
\end{tabular}
\caption{(Color online) Scaling combination $H$ defined in Eq.~(\ref{defH}) 
vs $\xi/L$ for the Gaussian (G) model and the $\pm J$ model at
$p=0.8$.  
}
\label{FSSH}
\end{figure}

The overlap magnetization $m$ and susceptibility are expected to behave
as
\begin{eqnarray}
m = \xi^{-\beta/\nu} u_h(T) f_m(\xi/L),
\label{scalingm}
\end{eqnarray}
and
\begin{eqnarray}
\chi = \xi^{2-\eta} u_h(T)^2 f_\chi(\xi/L),
\label{scalingchi}
\end{eqnarray}
where $f_m(x)$ and $f_\chi(x)$ are universal functions apart from a
multiplicative constant, and the scaling field $u_h(T)$ is an analytic
function of $T$.
If hyperscaling holds, we should have
\begin{eqnarray} 
\beta = \frac{\eta\nu}{2} .
 \label{eta-beta}
\end{eqnarray}
Thus, the combination
\begin{eqnarray}
    H = \frac{\xi^2 m^2}{\chi}
\label{defH}
\end{eqnarray}
should be a universal function of $\xi/L$, independent of the scaling
field $u_h(T)$.  In Fig.~\ref{FSSH}, we show the combination $H$ for the
Gaussian model and the $\pm J$ model at $p=0.8$.  The scaling is
good. Deviations are only observed for $\xi/L \lesssim 0.2$ --- these
data correspond to large temperatures --- and for $\xi/L \gtrsim 0.6$,
which, as discussed in Ref.~\cite{PPV-10}, is the region in which 
strong crossover effects are observed 
for the lattice sizes considered in this paper.
However, the observed trends are consistent with a unique
universal curve. We thus confirm the validity of Eq.~(\ref{eta-beta}),
independently of what the numerical value of $\eta$ is. If, indeed, 
$\eta = 0$, as theoretically predicted, Eq.~(\ref{eta-beta}) gives
$\beta = 0$.

\begin{figure}
\begin{tabular}{cc}
\includegraphics[width=16em,keepaspectratio]{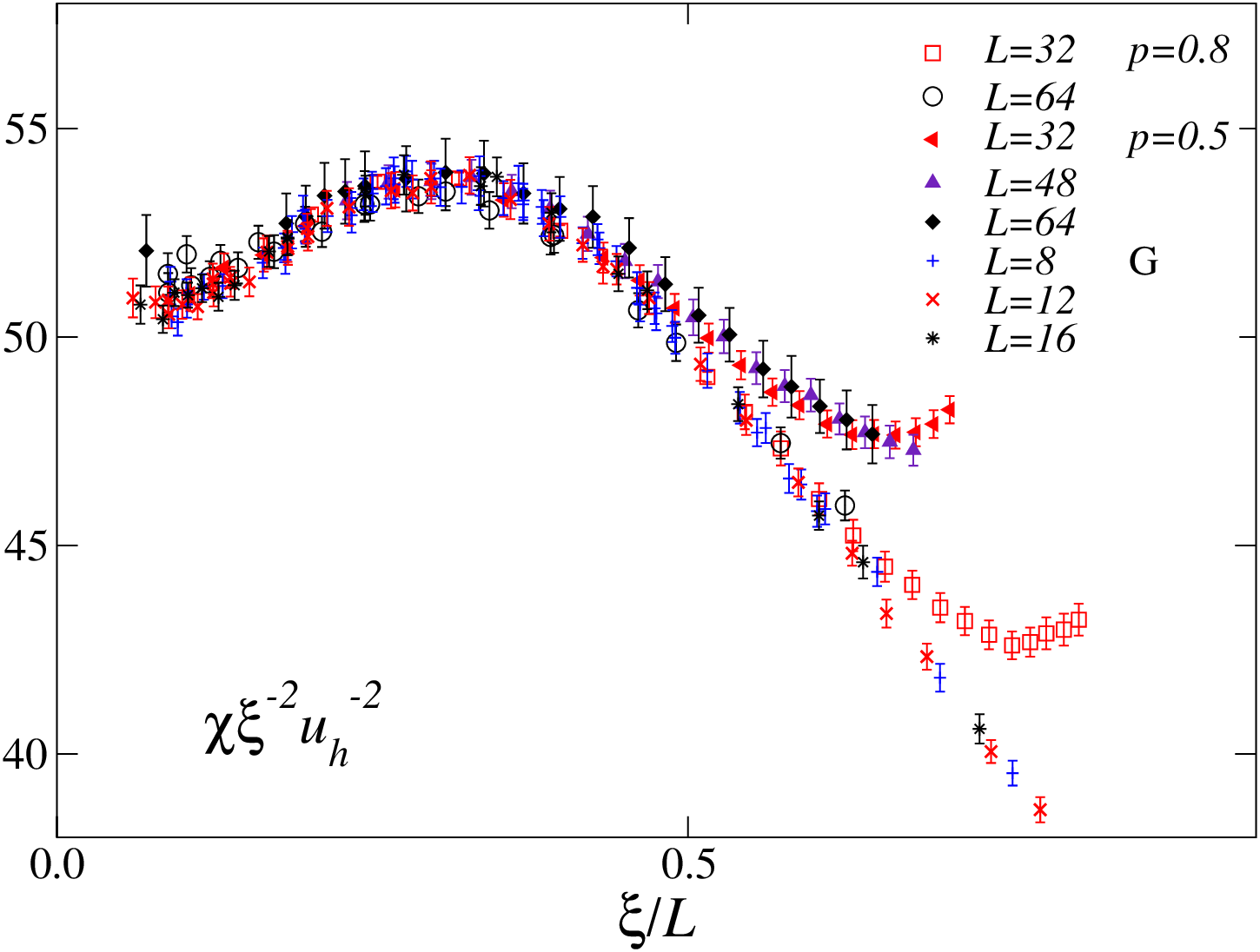} &
\hspace{1truecm} 
\includegraphics[width=16em,keepaspectratio]{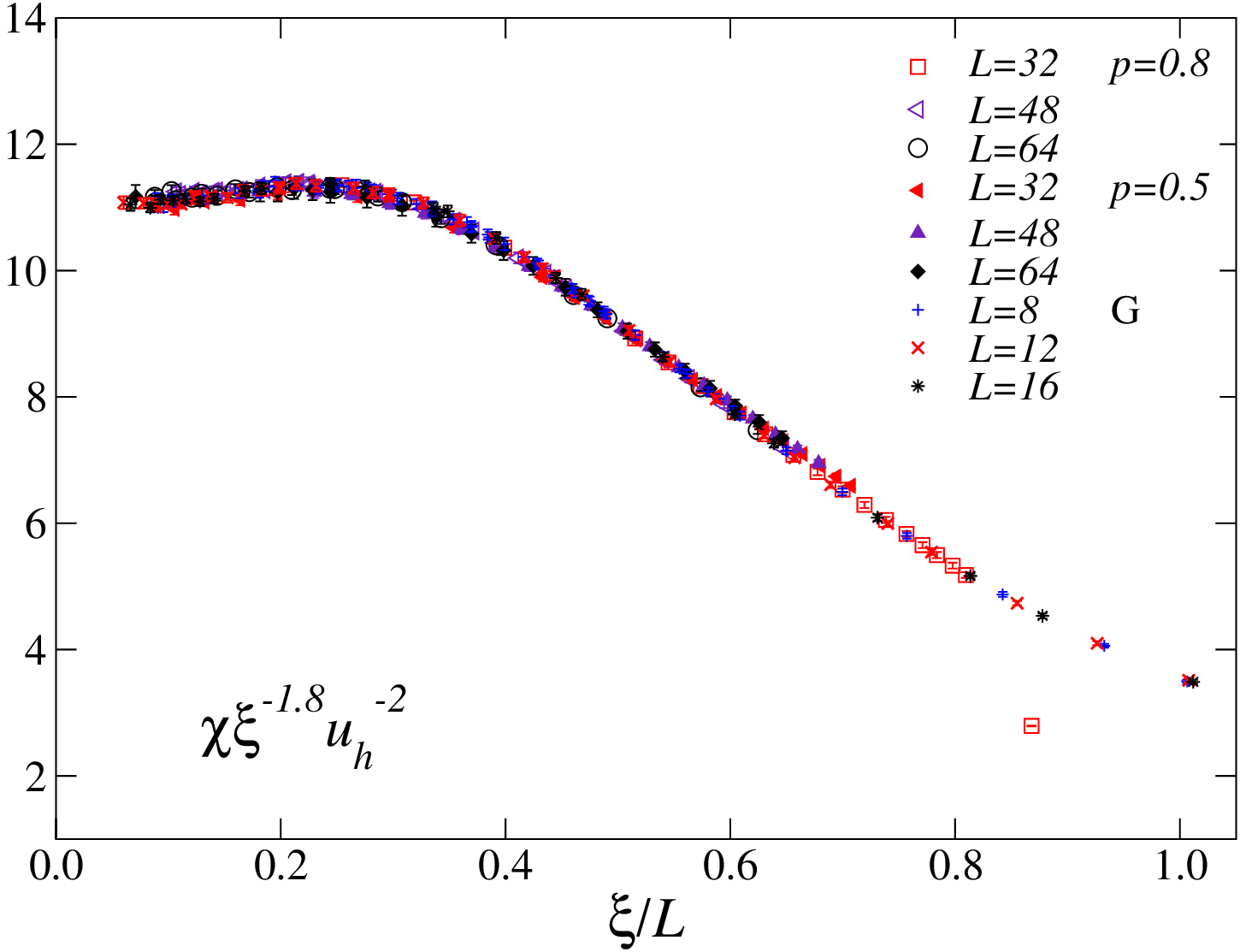} \\
\end{tabular}
\caption{(Color online) Scaling combination $\chi L^{\eta-2} u_h^{-2}$
vs $\xi/L$. We report the results for $\eta = 0$ (left) and $\eta
= 0.2$ (right).  Data are for the Gaussian model (G) and for the $\pm J$
model with $p=0.8$ and $0.5$.  
}
\label{FSSchi}
\end{figure}

We have fitted all Gaussian data for the overlap susceptibility to
Eq.~(\ref{scalingchi}) --- more precisely to its logarithm as in
Ref.~\cite{PPV-10} --- obtaining $\eta=0.20(7)$.  The error we report
is purely statistical and does not take into account possible scaling
corrections. This result is slightly larger than the predicted result
$\eta = 0$. The discrepancy should not be taken seriously, given the
small lattices we consider.  A precise determination of $\eta$ in the
Gaussian model requires, indeed, much larger values of $L$; see
Ref.~\cite{KLC-07}.

The function $f_\chi(x)$ is universal apart from a rescaling: if
$f_1(x)$ and $f_2(x)$ are determined in two different models, we expect
$f_1(x) = b f_2(x)$, where $b$ is a model-dependent constant. We now
compare the estimates of the functions $f_\chi(x)$ for the Gaussian
model and the $\pm J$ model. In Fig.~\ref{FSSchi}, we report the
functions (they have been rescaled so that they coincide for $\xi/L =
0.3$) for the Gaussian model and the $\pm J$ model at $p=0.8$ and $0.5$. We
report the curves both for $\eta = 0$ and $0.2$. In
Ref.~\cite{PPV-10}, we observed that for $\eta = 0$, the scaling was
good up to $\xi/L\approx (\xi/L)_{\rm max}$, a value which had been
estimated as the boundary of the crossover region before the regime in
which freezing was observed. The quantity $(\xi/L)_{\rm max}$ 
should scale as $(\xi/L)_{c}$, 
i.e., the value of $\xi/L$ at the crossover temperature.
Now, using Eq.~(\ref{RFSS}), since $h_{\xi/L}(x) \sim x^{-\nu}$
for $x\to 0$ to recover the correct infinite-volume behavior, we have
\begin{equation}
(\xi/L)_{\rm max} \sim (\xi/L)_{c} \sim 
   h_{\xi/L}[ a T_c(L) L^{1/\nu}] \sim T_c(L)^{-\nu}/L \sim 
       L^{\nu\theta_S - 1} \sim L^{0.8},
\end{equation}
where we have used $\theta_S = 0.5$ in the last step, and the fact
that $T_c(L) L^{1/\nu}\to 0$ for $L\to \infty$.
Again note that the inequality $\theta_S > 1/\nu$ is necessary to 
guarantee that $(\xi/L)_{\rm max} \to \infty$ as $L\to \infty$.
By looking at the scaling behavior of
$U_{22}$, Ref.~\cite{PPV-10} estimated $(\xi/L)_{\rm max} \approx
0.65$ and $0.45$ for $p=0.8$ and $0.5$, respectively, in the range $32\le L
\le 64$. For the $\pm J$ model at $p=0.5$, a similar estimate is
obtained by considering the scaling behavior of the estimates of
$\xi(2L)/\xi(L)$ reported in Ref.~\cite{JLMM-06}.  
The data for the Gaussian model agree with the $\pm J$
model data for both $p=0.5$ and $0.8$ up to $(\xi/L)_{\rm max}$.  
Thus, the numerical results are
consistent with universality and $\eta = 0$.

In Ref.~\cite{PPV-10}, we also observed that if we included all data in
the fits, the best estimate of $\eta$ was $\eta \approx 0.2$.  Indeed,
in this case, the results for the overlap susceptibility showed a very
good scaling up to $\xi/L\approx 0.8$.  We did not take this result as
an indication that $\eta = 0.2$ was a more plausible estimate than
$\eta = 0$ because we had good reasons to discard all data beyond
$\xi/L\approx (\xi/L)_{\rm max}$.  Somewhat surprisingly, if we now
include the data for the Gaussian model, and take $\eta = 0.2$ for all
models, we again observe a good universal scaling up to $\xi/L\approx
0.8$.  However, the numerical results of Ref.~\cite{KLC-07} exclude
$\eta=0.2$ for the Gaussian model. Thus, the apparently good observed
behavior cannot hold asymptotically.
In view of the result $\eta=0$ of Ref.~\cite{KLC-07}, as $L$ increases, the Gaussian data should become inconsistent with $\eta = 0.2$.

Finally, we note that,  in order to observe a good scaling behavior for
the overlap susceptibility, it is crucial to include the nonlinear
scaling field $u_h(T)$. Indeed, such a function gives a sizable
contribution to our data. In the case in which we set $\eta = 0$, we
obtain $u_h(T=1)^2/u_h(T=0.2)^2 \approx 2.2$ and
$u_h(T=1.5)^2/u_h(T=0.2)^2 \approx 3.3$.

\subsection{Two-point function} \label{sec3.3}

In Ref.~\cite{THM-11b}, the authors analyzed the scaling behavior 
of the two-point function, showing that below the crossover temperature, 
the two-point function scales as predicted by droplet theory. 
Here we analyze the two-point function in the opposite regime in which 
we expect 
\begin{equation}
   G_o(r,T) = {\xi^{-2} \chi} f_G(r/\xi,\xi/L).
\label{Gscaling}
\end{equation}
Note that by writing the scaling behavior in this form, there is neither
the need to specify $\eta$ nor to introduce the nonlinear scaling fields.
Moreover, the function $f_G(x,y)$ is universal.

\begin{figure}
\begin{tabular}{cc}
\includegraphics[width=17em,keepaspectratio]{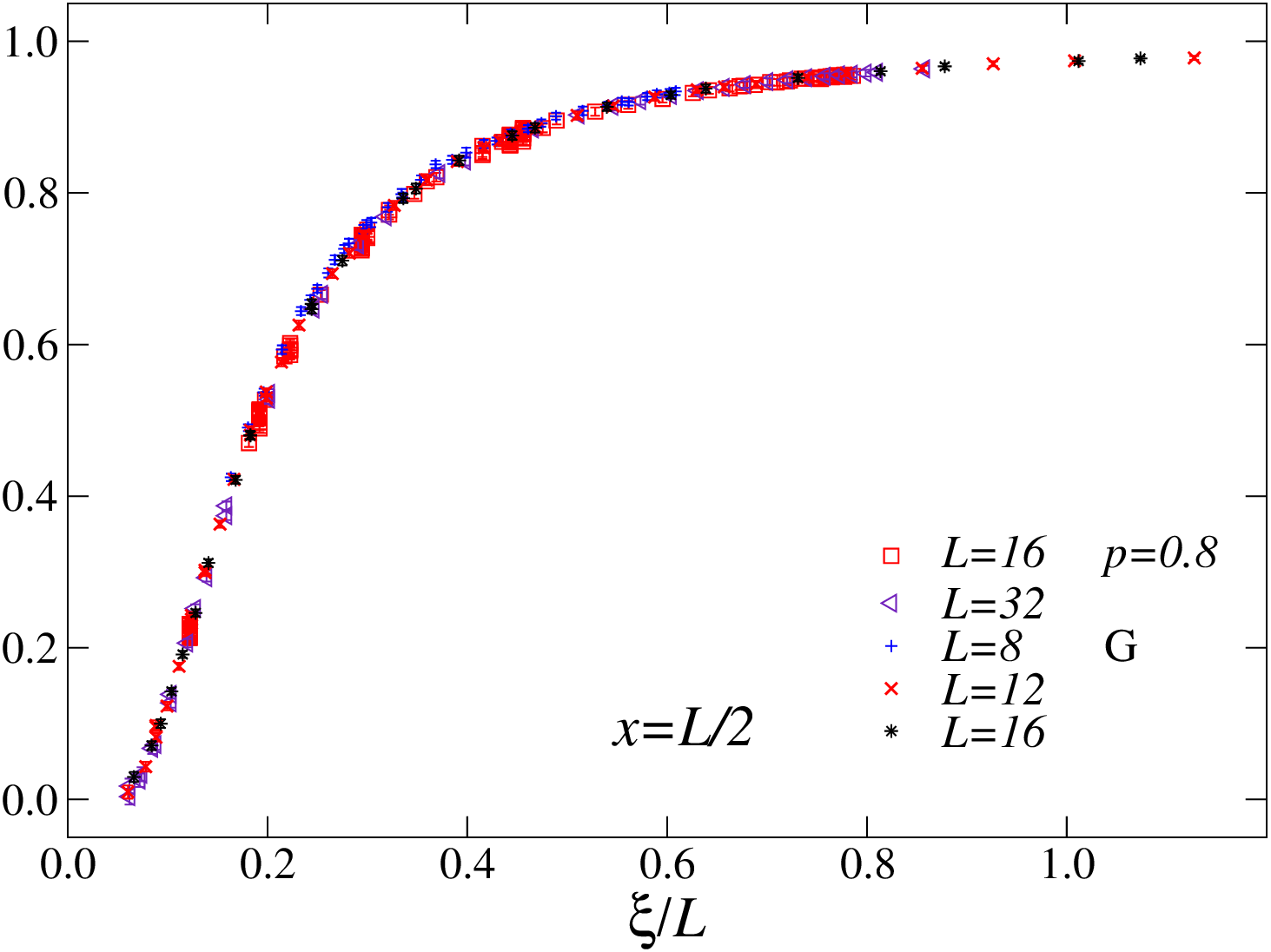} &
\hspace{1truecm}
\includegraphics[width=17em,keepaspectratio]{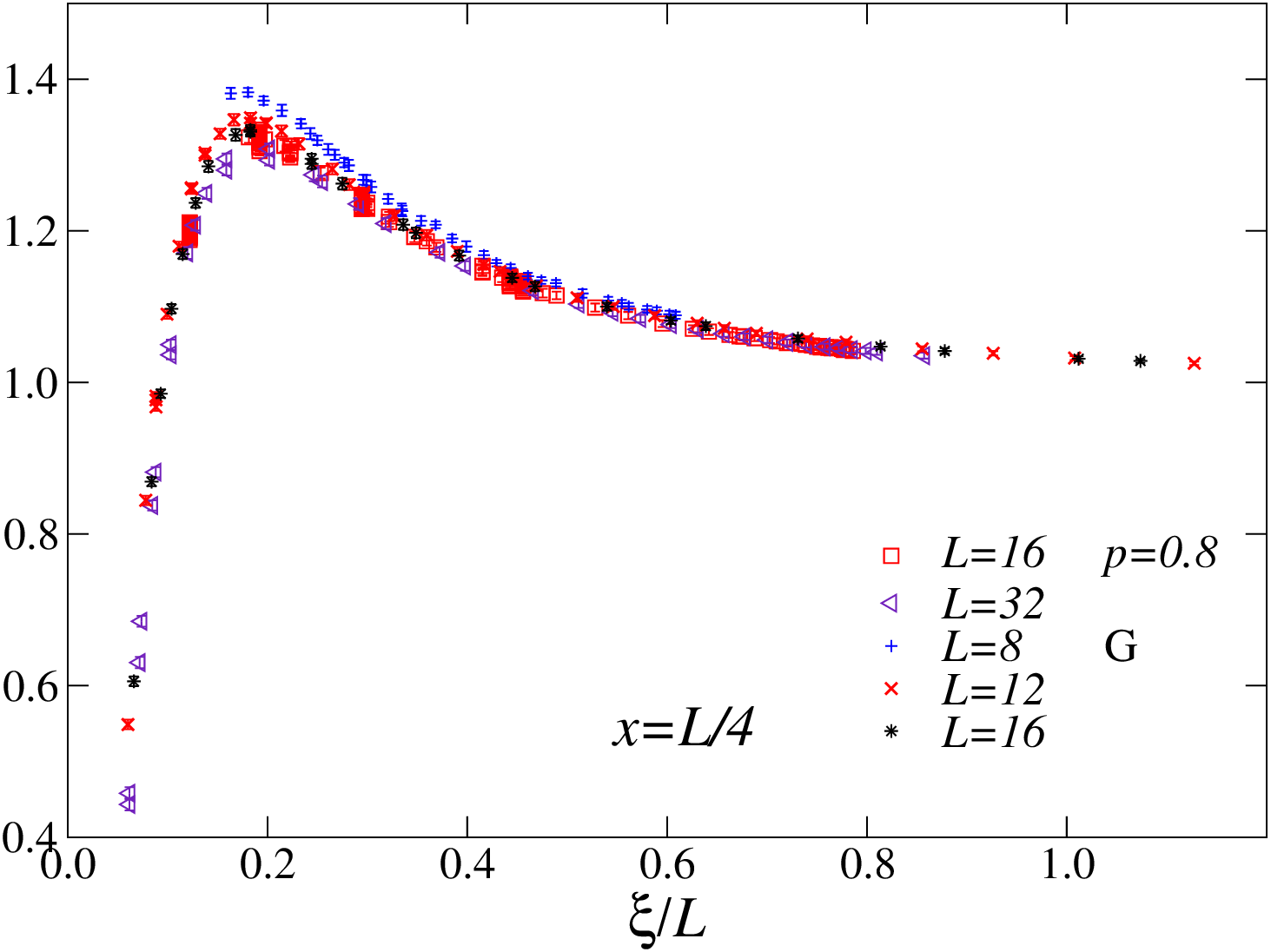} \\
\end{tabular}
\caption{(Color online) 
Scaling combination $g_o(x,T)$ for $x=L/2$ (left) and $x = L/4$ (right)
vs $\xi/L$. We report results for 
the Gaussian (G) model and the $\pm J$ model at $p=0.8$.
}
\label{FSSG_L}
\end{figure}

\begin{figure}
\begin{tabular}{cc}
\includegraphics[width=17em,keepaspectratio]{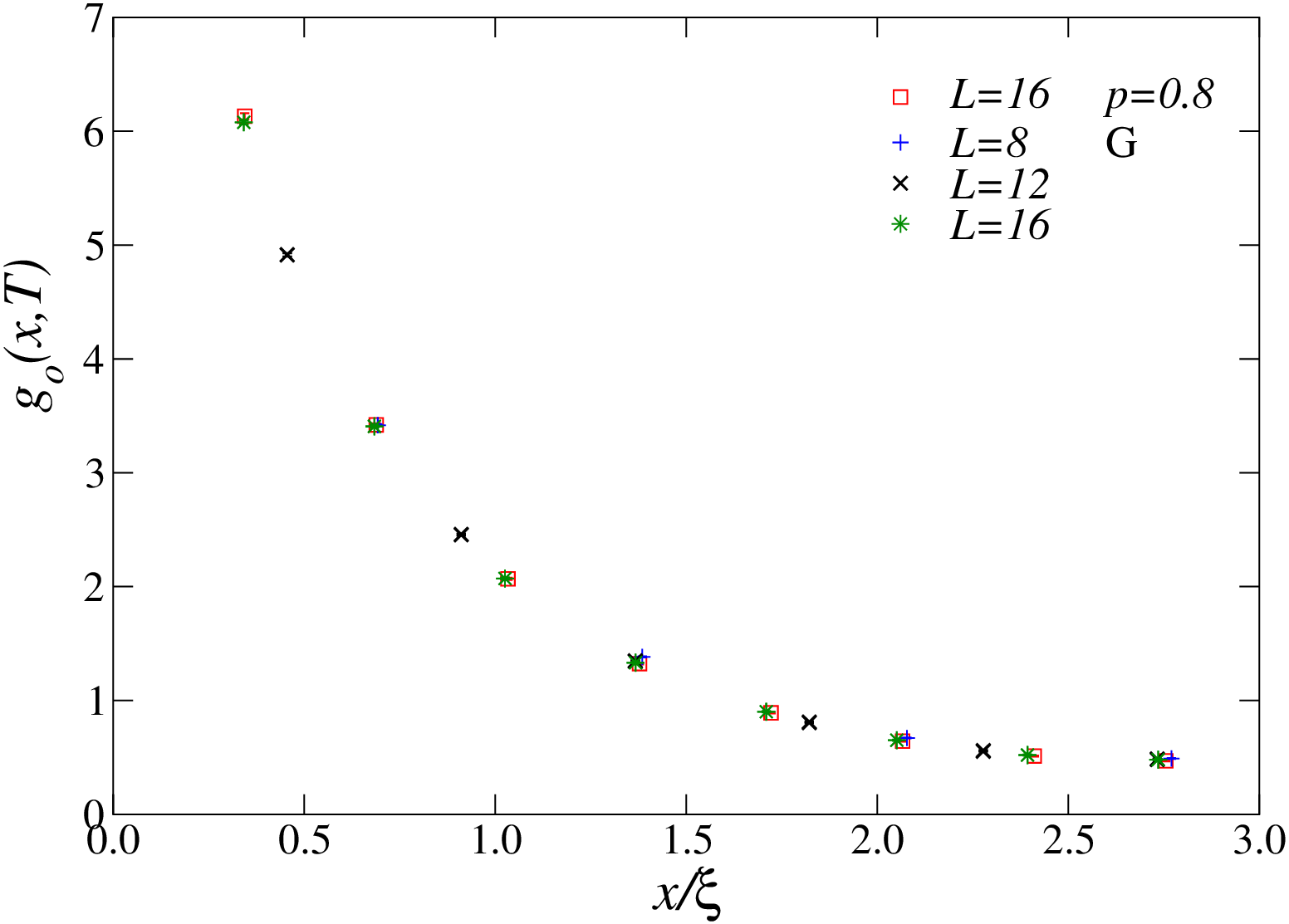} &
\hspace{1truecm}
\includegraphics[width=17em,keepaspectratio]{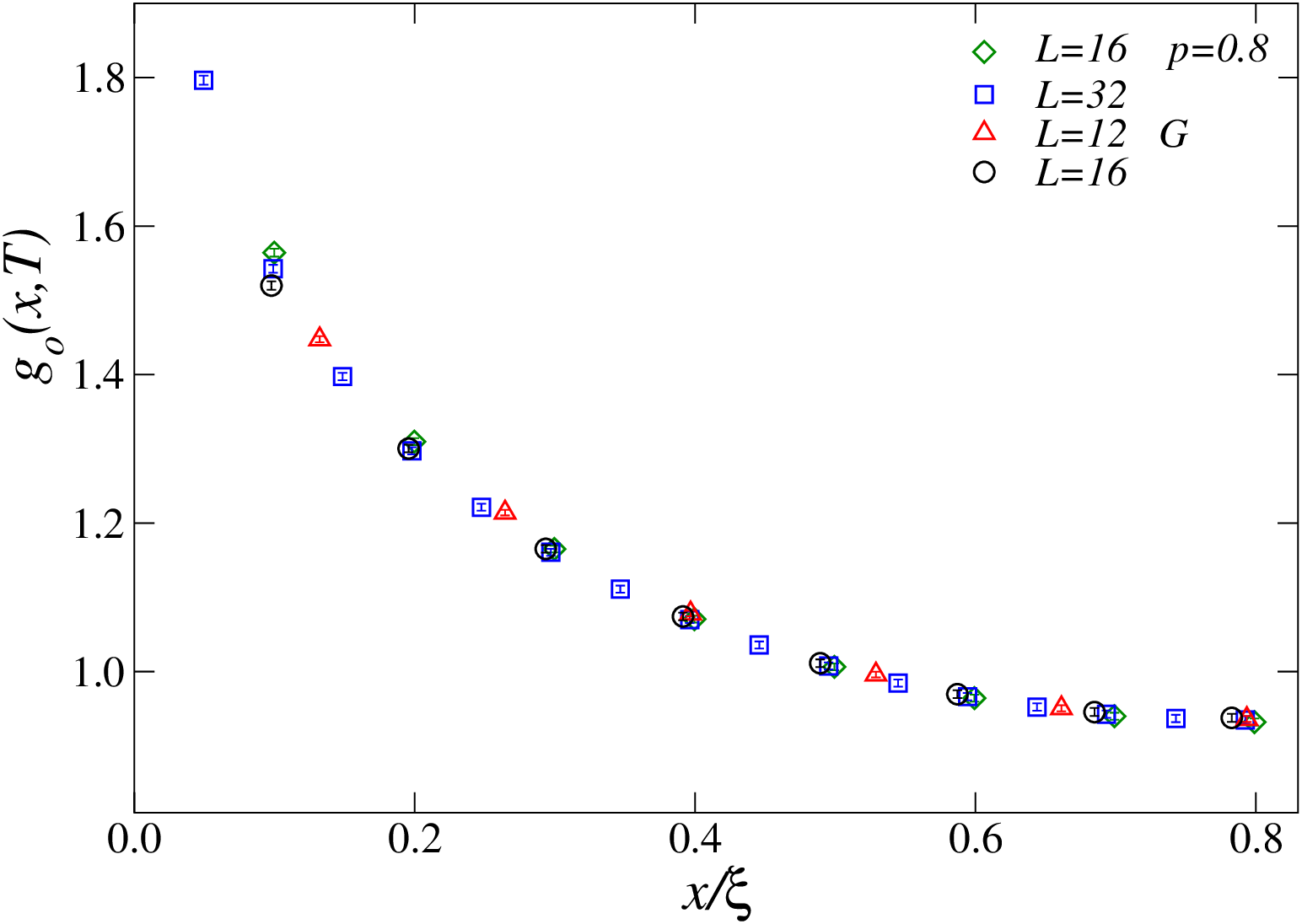} \\
\end{tabular}
\caption{(Color online) Scaling combination $g_o(x,T)$ for
$\xi/L\approx 0.18$ (left) and $\xi/L\approx 0.63$ (right) vs
$x/\xi$. We report the results for the Gaussian (G) model and the $\pm
J$ model at $p=0.8$.  }
\label{FSSG_x}
\end{figure}

To verify the scaling behavior (\ref{Gscaling}), we compute $G_o(r,T)$
along a lattice line, i.e. for $r = (x,0)$. We perform the computation
in the Gaussian model ($L=8,12,16$) and in the $\pm J$ model at
$p=0.8$ ($L=16,32$). Then, we consider
\begin{equation}
   g_o(x,T) \equiv \frac{G_o(r,T) L^2}{\chi}.
\end{equation}
In Fig.~\ref{FSSG_L}, we report $g_o(x,T)$ for $x=L/2$ and $L/4$ as a
function of $\xi/L$, while in Fig.~\ref{FSSG_x}, we show $g_o(x,T)$ at
fixed $\xi/L \approx 0.18$ and $\xi/L \approx 0.63$ as a function of
$x$. In these cases, the scaling is very good: all points fall onto a
single curve, confirming the validity of Eq.~(\ref{Gscaling}) and
universality.

\section{Conclusions} \label{sec4}

We investigate the FSS behavior of two-dimensional Ising spin-glass systems. In particular, we consider the square lattice $\pm J$
model at $p=0.5$ and 0.8, and the Gaussian model.  In this respect,
the $\pm J$ model appears particularly problematic because it
presents two different finite-volume regimes: a {\em continuous}
regime for $T>T_c(L)$ and a {\em discrete} regime for
$T<T_c(L)$. According to droplet theory, the crossover temperature
$T_c(L)$ is expected to vanish in the large-$L$ limit as a power
law~\cite{THM-11a,THM-11b,JK-11}, $T_c(L)\sim L^{-\theta_S}$ with
$\theta_S\approx 0.5$.  A logarithmic behavior, $T_c(L)\sim
1/\ln L$, is instead suggested by the free-energy arguments of
Ref.~\cite{KLC-07}.

The main conclusions of our numerical analysis based on MC simulations
are as follows:
\begin{itemize}
\item[(i)] All models we consider belong to the same universality
class.  The magnetization, susceptibility, two-point correlation
function, and the quartic cumulants, defined in terms of the overlap
variables, show a universal FSS behavior in terms of $\xi/L$.  In the
case of the $\pm J$ model, this universal scaling is only observed above 
the crossover temperature $T_c(L)$, which separates the {\em continuous} 
region from the {\em discrete} low-temperature behavior.
\item[(ii)] Our FSS analysis provides good evidence of the FSS
limit $T\to 0$, $L\to\infty$ at fixed $TL^{1/\nu}$.  This implies that
the crossover temperature $T_c(L)$ does not behave as $1/\ln
L$ as suggested in Ref.~\cite{KLC-07}, but rather as $T_c(L)\sim
L^{-\theta_S}$ with $\theta_S > 1/\nu \approx 0.28$. This is 
consistent with droplet theory, which predicts a 
power-law behavior with $\theta_S\approx 0.5$.
 \item[(iii)] We study the FSS behavior of $\chi$.
 The data for the Gaussian and the $\pm J$ models support universality.
 However, the available data are not sufficient to 
 obtain a precise estimate of $\eta$ and confirm definitely 
 the expected value $\eta=0$.
 \item[(iv)] We verify the hyperscaling relation $2\beta = \eta \nu$.
 If  $\beta=0$, it implies $\eta = 0$.
\item[(v)] We consider the two-point correlation function and show
  that it satisfies a standard FSS ansatz. Note that the scaling form
  (\ref{Gscaling}),
  which is appropriate for a high-temperature phase, is different from
  that considered in Ref.~\cite{THM-11b}, which is appropriate for a
  low-temperature phase in which spontaneous magnetization is present.
  The occurrence of these two different scaling behaviors is related to
  the different regimes considered. Here, we use data such that
  $T>T_c(L)$, while Ref.~\cite{THM-11b} studies the behavior for
  $T<T_c(L)$.
\end{itemize}


\begin{thebibliography}{99}
\bibitem{footnote-theta} We mention some of the most accurate
estimates of the stiffness exponent $\theta$ at $T=0$ in CES models:
$\theta=-0.281(2)$ (Ref.~\cite{RSBDJR-96}), $\theta=-0.282(2)$
(Ref.~\cite{HY-01}), $\theta=-0.282(3)$ (Ref.~\cite{CBM-02}), and
$\theta=-0.282(4)$ (Ref.~\cite{AMMP-03}) obtained by using the Ising
glass model with a Gaussian distribution for the couplings, and
$\theta=-0.275(5)$ (Ref.~\cite{LLMPV-07}) obtained in the
random-anisotropy model in the strong-anisotropy limit, whose glassy
critical behavior is in the same universality class.
\bibitem{logw} In Ref.~\cite{PPV-10}, the estimates of $T_f(L)$ were
only compared with a logarithmic behavior, finding a reasonable
consistency.  Although the statistical analysis of the data favors a
power-law behavior with $\theta_S\approx 0.35$, the available data
do not really allow us to conclusively discriminate between a
logarithmic and power-law approach to $T=0$.
\bibitem{footnote-etf} The error due to the fact that we are
considering only a discrete set of temperatures varies between 0.01
and 0.04, depending on $L$ and $p$. The estimate $T_f(L)$ also
slightly depends on which observable one considers.
\end{thebibliography}
\end{document}